\documentclass[12pt,a4paper]{article}
\usepackage{amsmath}
\usepackage{graphicx,psfrag,epsf}
\usepackage{enumerate}
\usepackage{natbib}
\usepackage{url} 

\newcommand{\blind}{0}

\usepackage{multirow,subcaption,amsfonts,stmaryrd,amssymb,amsthm,mathtools,bm,booktabs}
\usepackage{XCharter}
\usepackage[dvipsnames]{xcolor}
\usepackage[ruled]{algorithm2e}
\usepackage[backref=page]{hyperref}
\hypersetup{
    colorlinks=true,
    citecolor=blue,
}
\newcommand*{\cond}{\!\mid\!}
\newcommand*{\dquote}[1]{``{#1}"}
\newcommand*{\tr}{\mathsf{T}}
\newcommand*{\indic}[1]{\mathbb{I}{\set{{#1}}}}
\newcommand*{\ind}{\stackrel{ind}{\sim}}

\newcommand*{\set}[1]{\left\{{#1}\right\}}
\newcommand*{\bracket}[1]{\left({#1}\right)}

\begin{document}

\def\spacingset#1{\renewcommand{\baselinestretch}%
{#1}\small\normalsize} \spacingset{1}


\if0\blind
{
  \title{\bf Zero-inflated stochastic volatility model for disaggregated inflation data with exact zeros}
  \author{Geonhee Han\thanks{7-3-1 Hongo, Bunkyo-ku, Tokyo, 113-0033, Japan. gh2610@columbia.edu.}\hspace{.2cm}\\
    \small Graduate School of Public Policy, The University of Tokyo
    \\ \\
    Kaoru Irie\thanks{7-3-1 Hongo, Bunkyo-ku, Tokyo, 113-0033, Japan. irie@e.u-tokyo.ac.jp} \\
    \small Faculty of Economics, The University of Tokyo
    }
  \date{1 Mar 2026}
  \maketitle
} \fi

\if1\blind
{
  \bigskip
  \bigskip
  \bigskip
  \begin{center}
    {\LARGE\bf Zero-inflated stochastic volatility model for disaggregated inflation data with exact zeros}
\end{center}
  \medskip
} \fi

\bigskip
\begin{abstract}

The disaggregated time-series for the Consumer Price Index (CPI) often exhibits exact zero price changes, stemming from structural features of the {data collection process}.
However, the currently prominent stochastic volatility model of trend-inflation is designed for aggregate measures of price inflation, where zeros rarely occur.
We formulate a zero-inflated stochastic volatility model applicable to such non-stationary, real-valued, multivariate time-series data with exact zeros, which jointly specifies the dynamic zero-generating process.
For posterior inference, an efficient custom P\'{o}lya--Gamma augmented Gibbs sampler is derived.
Applying the model to disaggregated CPI data in four advanced economies --- US, UK, Germany, and Japan ---
we find that the zero-inflated model yields more informative estimates of time-varying trend and volatility, as it accounts for the presence of zeros and avoids underestimation.
In an out-of-sample forecasting exercise, we find that the zero-inflated model delivers improved point forecasts and better calibrated interval forecasts, particularly when zero-inflation is prevalent.

\end{abstract}

\noindent%
\vfill

\newpage
\spacingset{1.1}
\renewcommand{\arraystretch}{0.6}

\section{Introduction}

\paragraph{Background.}

In time series analysis, data are often transformed into first differences or rates of change to facilitate statistical modeling, inference, and prediction.
A classic example is the Box--Jenkins framework \citep[Chap.~9]{BoxJenkinsReinselLjung2015}, where trend and seasonal variations are removed by differencing.
The transformed time-series are then typically analyzed as continuous random variables, using established methodologies such as {dynamic linear models (DLMs)} and stochastic volatility (SV) models \citep[Chaps.~4 and 7]{West1997}.

Real-world time series data often exhibit instances in which the original values remain unchanged, resulting in exact zeros in the first differences or rates of change \citep{HautschMalecSchienle2013, KommKusters2015, GenaroSteffen2020}.
These frequently arise in economic datasets.
Stocks with low trading volume often experience price stasis, leading to zero returns.
Real estate investment trust (a.k.a., REIT) indices may exhibit zeros due to unobserved data points that are imputed using the most recent available value.
Vanilla state-space models, which assume continuous response (e.g., normal DLMs), can be unsuitable in the presence of such zeros, as location estimates may be biased toward zero and (conditional) variance may be underestimated, which is particularly problematic in applications where accurate estimation of time-varying means or volatilities is the fundamental task.

\paragraph{Motivation.}

This work has been motivated by the fact that the very phenomenon of zero-inflation is observed in lower-level disaggregated consumer price index (CPI) inflation data.
In the subsequent section, we illustrate using CPI data from four countries --- the United States (US), United Kingdom (UK), Germany (DE), and Japan (JP) --- that factors such as data collection practices contribute to price staleness.
\cite{KommKusters2015} also documents the frequent exact zeros in weekly price changes for skimmed whey powder in Germany, attributed to censoring and insufficient information.
Macroeconomic literature has further highlighted time- and state-dependent firm-level price adjustment behaviors to explain variations in the timing of price changes within firms \citep{KlnowKryvtsov2008, NakamuraSteinsson2008, DixonGrimme2022}, which confirm price adjustment behaviors at a sub-monthly frequency.

\paragraph{Related works.}

Despite zero-inflation, prominent non-stationary SV models designed for analyzing trend-inflation and volatility have largely overlooked the phenomenon of price staleness.
This is because these models were developed primarily to capture the distinctive characteristics of \textit{aggregate} inflation data, where the price staleness is dampened.
Namely, the seminal unobserved components model with stochastic volatility \citep[UCSV:][]{StockWatson2007} has become a widely adopted framework for modeling and forecasting aggregate inflation \citep{FaustWright2013}, with several methodological extensions proposed to refine its application to aggregate measures \citep[e.g.,][]{Chan2013, ChanKoopPotter2013, ChanKoopPotter2016, Chan2017, HwuKim2019, ZhangChanCross2020, HuberPfarrhofer2021}.

While there also exist models that account for zero-inflation or disaggregation, they are generally designed solely for predictive purposes and lack the capability to estimate trends and volatilities.
For example, \cite{KommKusters2015} used a univariate ARMA-GARCH model with a threshold mechanism for zeros.
\cite{BarkanBenchimolCaspiCohenHammerKoenigstein2023} employed hierarchical neural networks for disaggregated CPI forecasting.
\cite{PowellNasonElliottMayhewDaviesWinton2018} used autoregressive models to forecast disaggregated CPI based on monthly data and daily web-scraped prices.
The ability to estimate trends and volatilities with UCSV-type models is crucial, as they are fundamental features of macroeconomic time-series data \citep{Primiceri2005, Nakajima2011}, and serve as important information for policymakers and financial institutions in formulating financing strategies.

The alternative existing Bayesian approach is a set of models for censored time-series such as (dynamic) Tobit models \citep{Chib1992, Wei1999, LiZheng2008, LiuMoonSchorfheide2023}.
Here, the uni-directional censoring assumption is unsuitable for price inflation, which can take both positive and negative values, and more broadly zero-inflated continuous time series over the real line.
The latent threshold approaches \citep{NakajimaWest2013JBES, NakajimaWest2013JFE} may be appropriate if rounding is determined to be the sole cause of price stagnation.
This is false, as we shall demonstrate later.

There is also the choice of using discrete-valued sampling distributions. Such approaches are commonly found in the financial econometrics literature, where there is interest in modeling zero-inflation in essentially-discrete price movements in high-frequency financial transaction data \citep{HausmanLoMacKinlay1992, RydbergShephard2003, BienNoltePohlmeier2011}. In our specific case, the challenge requires simultaneously accommodating continuous observations and zero-inflation.

\paragraph{Scope.}

Our goal is to develop a zero-inflated multivariate SV model tailored for analyzing disaggregated zero-inflated price inflation data.
We generalize the (UC)SV model by explicitly specifying the exact-zero-generating process jointly with the (UC)SV-driven non-zero generating process, as a dynamic generalized linear model (DGLM) (\citealt{West1985a}; \citealt[Chap.~14]{West1997}).
In modeling the exact-zero-generating component, we incorporate the latent dynamic logistic model to allow for heterogeneous cross-sectional propensity, temporal persistence, and cross-correlation in zeros.

The idea of specifying the zero-generating process by using the dynamic logistic model can be seen in cases of discrete zero-inflated observations \citep[e.g.,][]{BerryWest2018DCMM, LavineCronWest2020factorDGLMs}, while our usage is for continuous response with SV.
Relatedly, \cite{YanchenkoDengLiCronWest2023} investigate zero-inflated household expenses by combining a dynamic logistic model for zeros and DLMs. While similar, our methodology is for continuous response over the real line, designed to additionally account for non-stationarities in variance.

In posterior computation, where the conditional posterior is intractable due to binomial likelihoods with a logistic link function, we utilize the P\'{o}lya--Gamma augmentation to restructure the intractable component as conditionally Gaussian and linear \citep{PolsonScottWindle2013}.
The augmentation enables us to construct a fast and efficient Gibbs sampler, where we can sample the state variables jointly \citep{WindleCarvalhoScottSun2013, glynn2019bayesian}.
This contrasts the approximate Bayesian approach to DGLMs \citep{West1985a} suitable for online forecasting for fast but approximate posterior computation, whereas our MCMC-based approach provides asymptotically exact inference for time-varying trends and volatilities (for our rather low-frequency data), which is crucial.

Now, the presented model and sampler are devised for modeling and forecasting CPI inflation data, but the flexible nature of DGLMs suggests wide applicability of our approach to a variety of real-valued multivariate time-series involving non-stationarity and zero-inflation.

\paragraph{Structure.}

The paper is structured as follows.
Section [\ref{section:2}] introduces the disaggregated CPI data and provides background on the occurrence of zeros.
Section [\ref{section:3}] presents the proposed zero-inflated (UC)SV, together with a brief overview of the posterior sampling algorithm.
Section [\ref{section:4}] applies the model to highlight the benefits of accounting for zero-inflation and the potential risks of ignoring zero-inflation.
Section [\ref{section:5}] reports the results of a forecasting exercise.
Section [\ref{section:6}] concludes the paper.

\section{Disaggregated consumer price indices} \label{section:2}

\subsection{Consumer price index} \label{sec:2:1}

The CPI serves as an important measure to track price fluctuations.
Specific implementations differ by country, but most national statistical agencies construct the CPI using a hierarchical bottom-up manner based on official survey data.

At the most granular level of the CPI hierarchy, prices are recorded for specific products in individual retail stores.
These observations are indexed by month, product, and store, which form \textit{time-item-location} level data.
These are then aggregated across locations (e.g., municipalities) by a weighted average to construct the \textit{time-item} level price indices.
These \textit{time-item} level indices feed into progressively higher levels of the CPI structure by averaging across lower-level categories, also using weights on their relative importance.
Such sequential aggregation yields a nested hierarchy of indices.
This is perhaps best exemplified by the \textit{Classification of Individual Consumption by Purpose} (COICOP) classification system used in many advanced economies:
overall aggregate index (COICOP-1),
divisions (COICOP-2),
groups (COICOP-3),
classes (COICOP-4),
and item-level indices (COICOP-5).
        
\subsection{Measurement error from data collection} \label{sec:2:2}

\begin{figure}[!h]
    \centering
    \includegraphics[width=\linewidth]{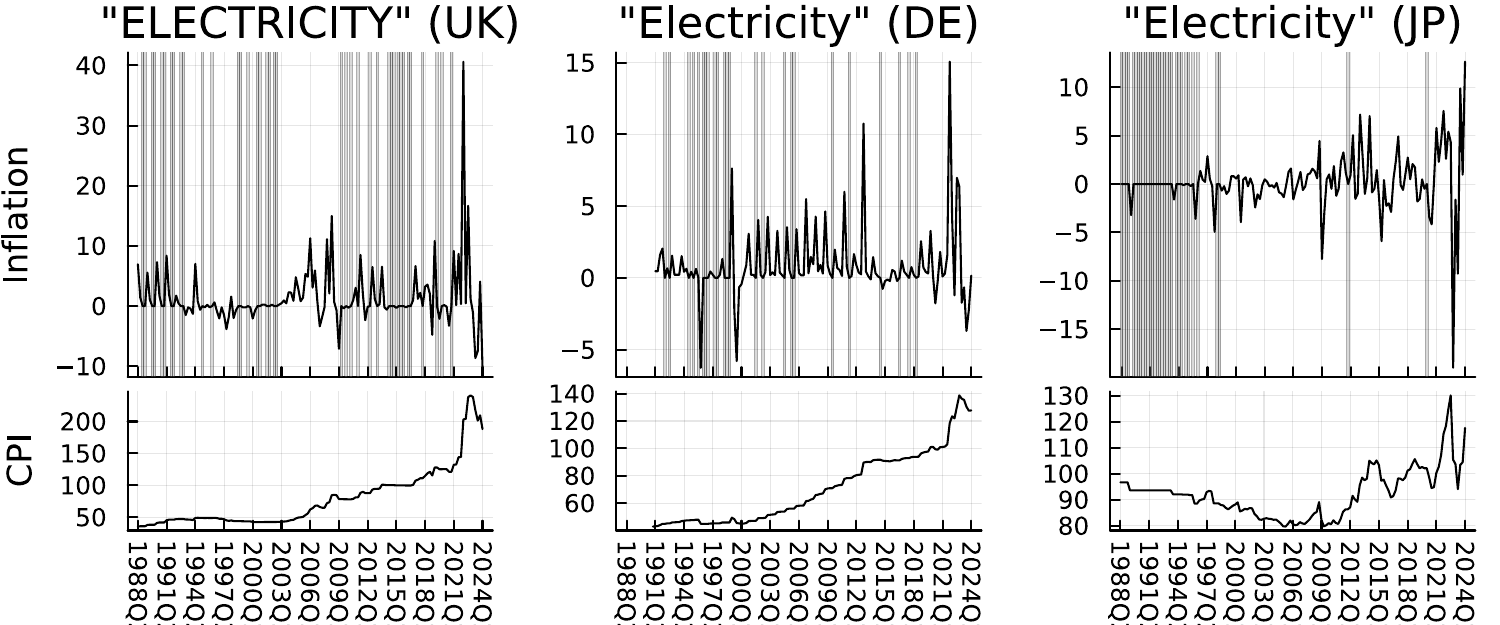}
    \spacingset{1.1} \caption{
        Time-series plot of the electricity component for three selected countries.
        The comparison highlights:
        (1) occurrences of exact zeros,
        (2) heterogeneity in the frequency and persistence of zeros,
        (3) non-stationarity in the non-zero component, and
        (4) missing values prior to component-level data collection.
        {\textbf{Definition}: Inflation (top) is defined as the quarterly \% change in CPI (bottom) from the previous quarter.}
        \textbf{Note}: Gray shaded vertical bands indicate periods of observed price staleness.
    }
    \label{fig:intro:data1}
\end{figure}

The hierarchical construction allows for capturing detailed price movements and, notably, reveals an important empirical feature;
exact zeros become increasingly common at finer levels of disaggregation.
Figure \ref{fig:intro:data1} visualizes one of these disaggregated components for the US, UK, Germany, and Japan.

The observed zero-inflation partly arises from the fact that, for each item-level index, national statistical agencies typically designate a fixed set of representative products --- often a singleton set --- for repeated price collection,
and \textit{month-product-store} data are often identified as \textit{time-item-location level} observations.
This design reflects a trade-off between precision and cost-efficiency;
tracking all prices for all predefined categories of \textit{item} is simply infeasible.
Even so, to best maintain consistency and comparability under finite data collection budgets, statistical agencies specify detailed product attributes beforehand, such as store identity, brand, package size, sales unit, production region, or model number.
The implication is that the observed price (staleness) reflects the behavior of a specific item from a particular firm at a given time and location, rather than the broader \textit{idealized} yet unavailable dynamics of the item and its aggregate category.

Moreover, some statistical agencies (e.g., BLS in the US and the SBJ) acknowledge that month-item-location specific indices may be missing, for instance when a product is discontinued in a surveyed municipality;
in such cases, the item and its associated weight may be excluded from higher-level index calculations.
Such is another similar layer of information scarcity.

Combined with the fact that many firms do not adjust prices monthly or even quarterly \citep{HigoSaita2007, DixonGrimme2022}, the scarcity of information contributes to the appearance of price staleness in disaggregated CPI data.
Our argument is therefore that the model should incorporate an additional layer of uncertainty to account for the discrepancy (i.e., a form of measurement error) between the observed data and the underlying price dynamics we aim to truly measure.

\subsection{{Missing values}} \label{sec:2:3}

\begin{figure}[!h]
    \centering
    \includegraphics[width=\linewidth]{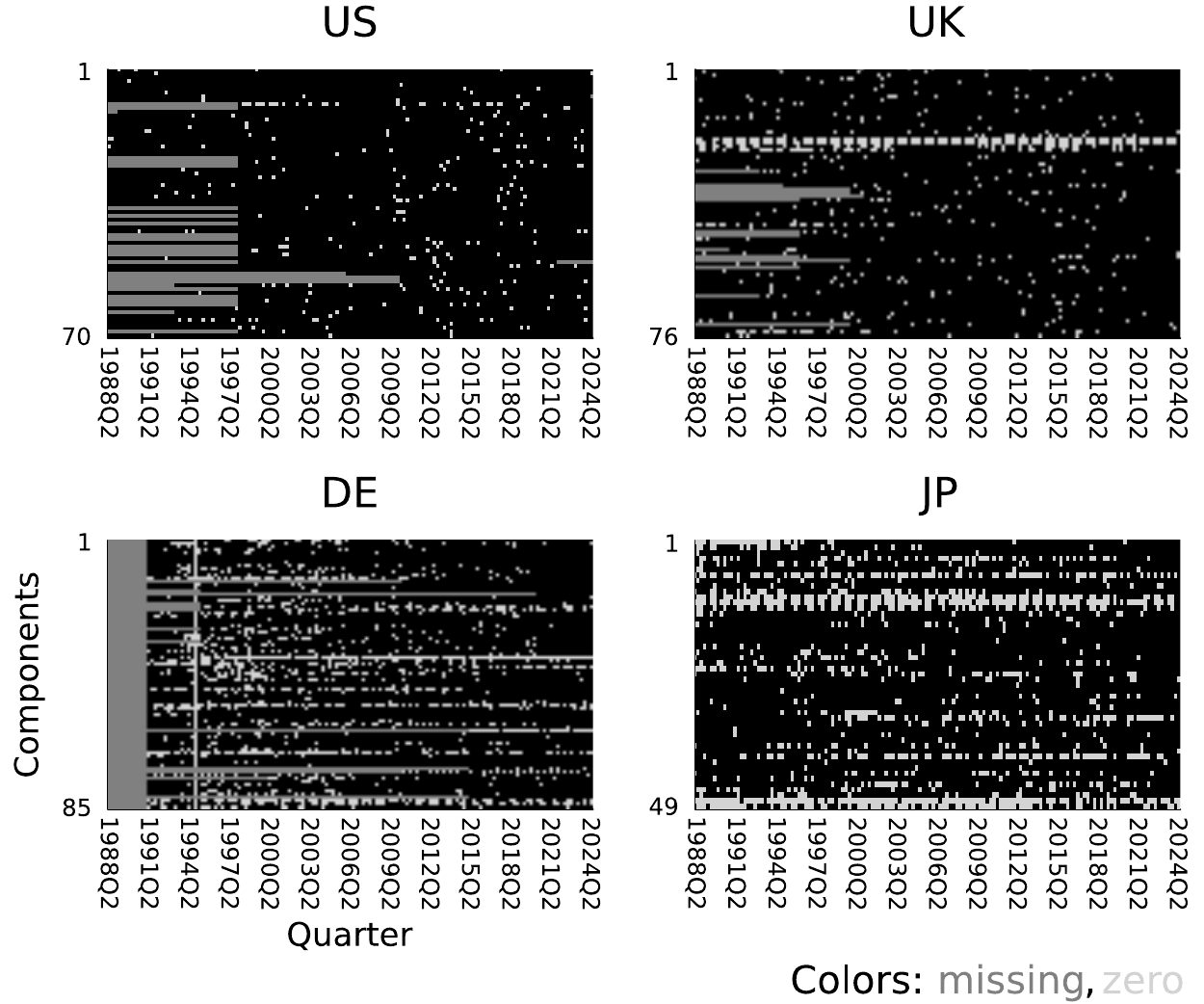}
    \spacingset{1.1} \caption{
        Heatmap of zero and missing-data patterns in the inflation time-series.
        zero-inflation and missing observations are prevalent features of disaggregated CPI inflation data, largely due to the mode of data collection.
    }
    \label{fig:intro:data2}
\end{figure}

Figure [\ref{fig:intro:data2}] illustrates the overall sparsity patterns in the disaggregated CPI data.
Although zero-inflation is our main focus, missing values are also not negligible.
The most notable reason for this is the variation in the start dates of data collection across items.
In addition, in some cases, items are surveyed less frequently than the standard monthly schedule, which contributes to intermittently missing observations.

\subsection{Intra- and inter-component dependence} \label{sec:2:4}

Figure \ref{fig:intro:data2} also highlights substantial variation in sparsity across components: with the proportion of exact zeros ranging from single-digit percentages to cases where non-zero observations are rare.
This cross-sectional heterogeneity in zero-inflation is another feature of the data and must be explicitly addressed in our model.

We also note strong inter-temporal persistence in zero-inflation, which we hypothesize to result from institutional factors.
A panel in Figure \ref{fig:intro:data1} illustrates this with quarterly inflation for Japanese electricity bills.
Prior to the 1995 revision of the \textit{Electricity Business Act} in Q4 of 1995 (which initiated energy liberalization in Japan), electricity prices were significantly stale with persistent zeros.
Such components are subject to price revisions only after administrative processes, which leads to observable inflation occurring intermittently and contributing to persistent zeros.
Similar patterns are evident in other regulated categories, such as medical services and school fees, which also exhibit high proportions of zeros.
Conversely, items whose prices are largely market-driven, such as food, are less prone to zero-inflation.

\begin{figure}[h]
    \centering
    \includegraphics[width=\linewidth]{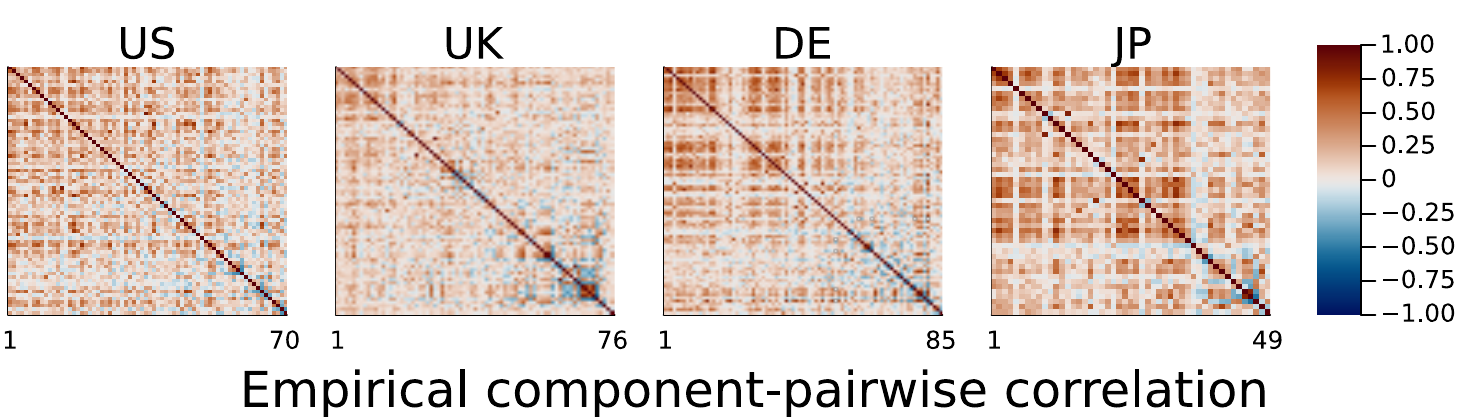}
    \spacingset{1.1} \caption{
        Heatmap of the empirical component-pairwise correlation matrix, including both zero and non-zero observations.
        The empirical correlation reinforces the necessity of modeling inter-component dependencies when estimating trend inflation and volatility.
    }
    \label{fig:intro:data5}
\end{figure}

Capturing within-component inter-temporal persistence of zeros (or its absence) would thus be another requirement for the model.
In doing so, especially zero-inflation, one should not only exploit intra-component persistence,
but also incorporate inter-component co-movements across disaggregates as summarized in Figure \ref{fig:intro:data5} to best characterize the underlying inflation dynamics.

\subsection{Summary}

We summarize the desirable features of the zero-inflated model as follows.
\begin{itemize}
    \item The most fundamental ability to estimate trends and SVs should be kept consistent with established models in the literature (e.g., \citealp{StockWatson2007, StockWatson2016, EoEzedaWong2023}).
    \item Upon the UCSV structure, the model should account for the heterogeneity in the propensity for zeros (Section \ref{sec:2:2}).
    \item For the zero and non-zero component, the model should capture temporal persistence, and potential cross-series dependencies in the (non-)zero-generating mechanism, given that the time series is multivariate (Section \ref{sec:2:3}).
    \item Missingness should also be addressed simultaneously (Section \ref{sec:2:4}).
    Following \cite{BerryWest2018DCMM}, it would be desirable to jointly model zero, non-zero, and missing observations within a joint probabilistic model.
\end{itemize}
In the following section, we present a SV model with zero-inflation that addresses these requirements.

\section{Zero-inflated UCSV model} \label{section:3}

\subsection{The model}

\paragraph{A state-space model of trend-inflation.}

We first propose a univariate formulation of the zero-inflated UCSV (Z-UCSV) model.
To this end, let $y_{1:T} := (y_1, ...., y_T)^\tr$ represent the univariate observations, and
$\theta_{1:T} := (\theta_1, ..., \theta_T)^\tr$
and $h_{1:T} := (h_1,...,h_T)^\tr$
be the unobserved trend and
measurement log-volatility.

The UCSV model we consider relates the triple $(y_{1:T}, \theta_{1:T}, h_{1:T})$ by
\begin{align}
    y_t &= \theta_t + \varepsilon_t^{(y)},
    &
    \varepsilon_t^{(y)} \cond h_t &\ind \textsc{N}_1(0, \exp{h_t}),
    \label{eq:ucsv:uni:y} \\
    \theta_t &= \theta_{t-1} + \varepsilon_t^{(\theta)},
    &
    \varepsilon_t^{(\theta)} \cond \sigma^2_{\theta} &\ind \textsc{N}_1(0, \sigma^2_{\theta}),
    \label{eq:model:uni:theta} \\
    h_t &= h_{t-1} + \varepsilon_t^{(h)},
    &
    \varepsilon_t^{(h)} \cond \sigma_{h}^2 &\ind \textsc{N}_1(0, \sigma_{h}^2),
    \label{eq:model:uni:h}
\end{align}
where $\textsc{N}_D(\bm{\mu}, \bm{\Sigma})$ is used to denote the $D$-variate normal distribution with mean and covariance $(\bm{\mu},\bm{\Sigma})$.
{The index runs through $t \in [T] := \set{1, \ldots, T}$.}
The variances $\sigma^2_{\theta}, \sigma^2_{h} > 0$ and initial states $\theta_0, h_0$ are also random, to be specified later with priors.

The time-varying conditional means $\theta_{1:T}$ are often referred to as a latent trend component or \textit{trend-inflation} \citep{StockWatson2007, Chan2013, ChanKoopPotter2013, ChanKoopPotter2016, StockWatson2016, LiKoopman2021, EoEzedaWong2023}, reflecting the belief that the underlying process governing the permanent level of inflation evolves smoothly.
Movements in the trend are interpreted as \textit{permanent} (potentially persistent) shifts in inflation.
The deviation from this trend $y_{1:T} - \theta_{1:T}$ is seen as \textit{transient} (temporary) shifts about the permanent level  \citep[a.k.a., \textit{inflation gap}]{CogleyPrimiceriSargent2010, HwuKim2019}.
The log-volatility process $h_{1:T}$ represents transitory volatility.
The original formulation also includes another log-volatility process in the transition equation, in place of time-invariant variance $\sigma^2_{\theta}$, which represents permanent volatility.
This component is omitted in our model, as variations in volatility are much less pronounced at the item-specific level;
similar implementations are prevalent in existing literature \citep[e.g.,][]{Chan2013}.

\paragraph{Zero-inflated model.}

We introduce three additional latent processes:
$y^*_{1:T} := (y_1^*, ..., y_T^*)^\tr$,
$p_{1:T} := (p_1,...,p_T)^\tr$,
and $\pi_{1:T} := (\pi_1,...,\pi_T)^\tr$.
We specify a latent dynamic logistic model, a special case of DGLM (\citealt{West1985a}; \citealt[Chap.~14]{West1997}) by
\begin{align}
    y_t \cond y_t^*, p_t &\ind (1 - p_t) \, \delta_{y_t^*} + p_t \, \delta_{0}, \label{eq:model:uni:y} \\
    y_t^* &= \theta_t + \varepsilon_t^{(y)}, \label{eq:model:uni:z} \\
    p_t &= \operatorname{logit}^{-1}(\pi_t), \label{eq:model:uni:p} \\
    \pi_t &= \pi_{t-1} + \varepsilon_t^{(\pi)},
    \qquad \qquad \qquad
    \varepsilon_t^{(\pi)} \cond \sigma^2_{\pi} \ind \textsc{N}_1(0, \sigma^2_{\pi}). \label{eq:model:uni:pi}
\end{align}
The processes $(\theta_{1:T}, \varepsilon_t^{(y)}, h_{1:T})$ inherit the same dynamical specification as in eq.~\eqref{eq:ucsv:uni:y}, \eqref{eq:model:uni:theta}, and \eqref{eq:model:uni:h}.
$\delta_{c}$ is the point-mass at $c$.
Note that, $y_{1:T}$ in eq.~\eqref{eq:model:uni:y} need not equate to the almost-surely non-zero $y_{1:T}^*$ \textit{a priori};
this reflects the discrepancy between what we observe and what we want to measure.
Also, the probabilities $p_{1:T}$ are granted persistence via a Markovian structure $p(\pi_{1:T} \cond \sigma^2_{\pi}, \pi_0) = \prod_{t=1}^T p(\pi_t \cond \pi_{t-1}, \sigma^2_{\pi})$.
{Both aspects are motivated by the desiderata outlined in Section \ref{section:2}.}

The model is completed with priors on the static parameters and initial states,
\begin{align}
    \sigma^2_{\theta} \sim \textsc{IG}(\alpha^{(\theta)},\beta^{(\theta)}), \qquad
    \sigma^2_{h} &\sim \textsc{IG}(\alpha^{(h)},\beta^{(h)}), \qquad
    \sigma^2_{\pi} \sim \textsc{IG}(\alpha^{(\pi)},\beta^{(\pi)}),
    \label{eq:model:uni:prior1} \\
    \theta_0 \sim \textsc{N}_1(m^{(\theta)}, v^{(\theta)}), \qquad
    h_0 &\sim \textsc{N}_1(m^{(h)}, v^{(h)}), \qquad
    \pi_0 \sim \textsc{N}_1(m^{(\pi)}, v^{(\pi)}),
    \label{eq:model:uni:prior2}
\end{align}
where $\textsc{IG}(\alpha, \beta)$ represents the inverse gamma distribution with density $\textsc{IG}(x \cond \alpha, \beta) \propto x^{-(\alpha + 1)} \exp(-\beta / x)$.

To summarize, the observed univariate time-series $y_{1:T}$ relates to the unobserved collection of interest,
\begin{equation}
    \bm{\Theta} := (
    y_{1:T}^*, \;
    \theta_{0:T}, \;
    h_{0:T}, \;
    \pi_{0:T}, \;
    \sigma^2_{\theta}, \;
    \sigma^2_{h}, \;
    \sigma^2_{\pi}
    ),
    \label{eq:model:uni:param}
\end{equation}
via the joint distribution with hyperparameters $(
    m^{(*)}, v^{(*)},
    \alpha^{(*)}, \beta^{(*)}
)$ over $* \in \set{\theta, h, \pi}$ with
\begin{align*}
    p(y_{1:T}, \bm{\Theta})
    &=
    p(\theta_0) \,
    p(h_0) \,
    p(\pi_0) \,
    p(\sigma^2_{\theta}) \,
    p(\sigma^2_{h}) \,
    p(\sigma^2_{\pi})
    \\
    & \quad \times \prod_{t=1}^T
    p(\theta_t \cond \theta_{t-1}, \sigma_{\theta}^2) \,
    p(h_t \cond h_{t-1}, \sigma^2_h) \,
    p(\pi_t \cond \pi_{t-1}, \sigma^2_{\pi}) \,
    p(y_t^* \cond \theta_t, h_t) \,
    \mathbb{P}(y_t \cond y_t^*, \pi_t).
\end{align*}
The UCSV model \citep{StockWatson2007} is a sub-model;
in the limit $\pi_t \rightarrow -\infty$, it holds $p_t = 0$, under which it also holds
\begin{equation}
    y_t = y_t^* = \theta_t + \varepsilon_t^{(y)}. \label{eq:model:uni:y=z}
\end{equation}

\subsection{Posterior inference}

\paragraph{Markov-chain Monte Carlo.}

An overview of a Gibbs sampling algorithm for posterior inference is outlined in Algorithm \ref{alg:1}.
A detailed exposition is provided in the appendix.

\begin{center}
\spacingset{1}
\begin{minipage}{\linewidth} 
\begin{algorithm}[H]
    \medskip
    
    \begin{enumerate}[(a)]
    \item Sample the initial states $(\theta_0, h_0, \pi_0)$ given $(\theta_1, h_1, \pi_1, \sigma^2_{\theta}, \sigma^2_h, \sigma^2_{\pi})$.
    
    \item Sample the static parameters $(\sigma^2_{\theta}, \sigma^2_h, \sigma^2_{\pi})$ given $(\theta_{0:T}, h_{0:T}, \pi_{0:T})$.

    \item Sample the latent trend $\theta_{1:T}$ given $(\theta_0, \sigma_{\theta}^2, y_{1:T}^*)$.
    
    \item Sample the stochastic volatility $h_{1:T}$ given $(h_0, \sigma^2_h, \theta_{1:T}, y_{1:T}^*)$.
    
    \item Sample the dynamic probabilities $(\operatorname{logit}(p_t) = \pi_t)_{t=1}^T$ given $(\pi_0, \sigma^2_{\pi}, y_{1:T})$.

    \item Data-augment non-zero $y_{1:T}^*$ given $(\theta_{1:T}, h_{1:T}, y_{1:T})$.
    \end{enumerate}

    
    \caption{\texttt{Gibbs sampler: univariate model}} 
    \label{alg:1}   
\end{algorithm}
\end{minipage}
\end{center}

In steps (a) and (b), the static parameters and initial states are readily simulated from their full conditionals.
In step (c), the full conditional of the latent trend is that of a DLM; sampling is doable using any apparatus to sample the latent states in a linear Gaussian state-space model.
We make use of the forward-filtering backward-sampling algorithm \citep[FFBS:][]{CarterKohn1994, Fruhwirthschnatter1994}.
In step (d), the log-volatility processes are efficiently simulatable via auxiliary normal mixtures \citep{KimShephardChib1998}.
We make use of the ten-component mixture approximation \citep{OmoriChibShephardNakajika2007}.
The states are then sampled jointly using precision samplers \citep{ChanJeliazkov2009} for their efficiency.
In step (f), we draw $y_t^*$ from eq.~\eqref{eq:model:uni:z} given $(\theta_t, h_t)$ at $t$ such that $y_t = 0$.

\paragraph{P\'{o}lya--Gamma augmented FFBS.}

Step (e) involves sampling from the full conditional posterior of the dynamic pre-transformed probabilities $p(\pi_{1:T} \cond \pi_0, \sigma^2_{\pi}, y_{1:T})$ with logistic transformations and binomial likelihoods, {then explicitly retrieving $p_t = \operatorname{logit}^{-1}(\pi_t)$ as in eq.~\eqref{eq:model:uni:p}}.
To do so, first introduce the auxiliary sparsity indicators $\gamma_{1:T} := (\gamma_1, \; ..., \; \gamma_T)^\tr$ where $\gamma_t \cond p_t \ind \textsc{Bi}(1, p_t)$ (binomial distribution with a single trial and success probability $p_t$),
and replace the measurement equation in eq.~\eqref{eq:model:uni:y} with $y_t = y_t^* \indic{\gamma_t = 0}$, where $\indic{\cdot}$ is an indicator function.
The \textit{likelihoods} on $\gamma_{1:T}$ are also re-written via augmented likelihoods on $(\gamma_t, \omega_t \cond \pi_t)$,
\begin{align*}
    \mathbb{P}(\gamma_{t} \cond \pi_{t})
    &= {(\exp{\pi_t})^{\gamma_t} \over 1 + \exp{\pi_t}}
    = \int_0^{\infty} p(\gamma_t, \omega_t \cond \pi_t) \; d\omega_t \\
    &= \int_0^{\infty}
        2^{-1}
            \exp\!\set{(\gamma_t - 1/2) \pi_t} \,
            \exp\!\set{-\omega_t \pi^2_t / 2}
            \textsc{PG}(\omega_t \cond 1, 0) \; d\omega_t
    .
    \tag{$t \in [T]$}
\end{align*}
$\textsc{PG}(\cdot \cond 1, 0)$ is the probability density of the P\'{o}lya--Gamma distribution $\textsc{PG}(1, 0)$ following the conventional notation \citep[see][]{PolsonScottWindle2013, WindleCarvalhoScottSun2013}. With a prior density $p(\pi_{1:T} \cond \pi_0, \sigma^2_{\pi})$, the joint distribution is locally re-expressed by
\begin{align*}
    p(\pi_{1:T}, \gamma_{1:T} \cond \pi_0, \sigma^2_{\pi})
    &= p(\pi_{1:T} \cond \pi_0, \sigma^2_{\pi}) \prod_{t=1}^T \int_0^{\infty} p(\gamma_t, \omega_t \cond \pi_t) \; d\omega_t \nonumber
    \\ &
    = \int_{(0, \infty)^T} p(\pi_{1:T} \cond \pi_0, \sigma^2_{\pi}) \, p(\gamma_{1:T}, \omega_{1:T} \cond \pi_{1:T}) \; d\omega_{1:T}.
\end{align*}

The integrand in the above characterizes a conditional joint distribution on $(\pi_{1:T}, \gamma_{1:T}, \omega_{1:T})$ that admits $p(\pi_{1:T}, \gamma_{1:T} \cond \pi_0, \sigma^2_{\pi})$ as its marginal.
This facilitates simple, efficient, and practical sampling from the target $p(\pi_{1:T} \cond \pi_0, \sigma^2_{\pi}, \gamma_{1:T}) \propto p(\pi_{1:T}, \gamma_{1:T} \cond \pi_0, \sigma^2_{\pi})$ via conditioning on $\omega_{1:T}$, performed in two steps using the following two densities.
\begin{align*}
    p(\omega_{1:T} \cond \pi_{1:T})
        &\propto \prod_{t=1}^T \textsc{PG}(\omega_t \cond 1, \pi_t), \\
    p(\pi_{1:T} \cond \pi_0, \sigma^2_{\pi}, \gamma_{1:T}, \omega_{1:T})
        &\propto
        \underbrace{p(\pi_{1:T} \cond \pi_0, \sigma^2_{\pi})}_{\substack{\text{Linear Gaussian} \\ \text{latent state}}} \,
        \underbrace{\prod_{t=1}^T \exp{\set{ -{\omega_t \over 2} \bracket{ \pi_t - {\gamma_t - 1/2 \over \omega_t}}^2 } }}_{\text{Independent normal \dquote{likelihoods}}}.
\end{align*}
Since the \textit{prior} was specified to be a conditionally linear and Gaussian in eq.~\eqref{eq:model:uni:pi}, we may simply rely on the FFBS algorithm for the second step.


\paragraph{Remarks.}

{The approach to specify a two‑/multi-point distribution with a (dynamic) binary/multinomial latent logistic model is common across a range of statistical applications,
including but not limited to
mixture modeling \citep[e.g.,][]{LindermanJohnsonAdams2015, RigonDurante2021} and
structured variable selection \citep{StingoChenVanucciBarrierMirkes2010, StingoChenTadesseVannucci2011}.
The strategy to pair (i) PG augmentation with (ii) an efficient sampler to draw from an appropriate linear normal distribution, applies in similar such settings whenever the local likelihood is binomial and the latent variables are normal.
As for (i), drawing PG variates is highly efficient in the sense that a uniform rejection-sampling acceptance probability no lower than $0.99919$ has been demonstrated \citep[pg.~1340]{PolsonScottWindle2013} based on the alternating series method \citep{Devroye1986}.
}


{We note that several sampling strategies are applicable for $(\pi_{1:T} \cond -)$, and justify our choice here.
We choose P\'{o}lya--Gamma augmentation because it is practical and computationally tractable in exactly targeting the full conditional distribution while avoiding prohibitive costs.
For example, approximate auxiliary mixture approaches \citep[e.g.,][]{FruhwirthSchnatterFruhwirth2007} are available, but P\'{o}lya--Gamma random sampling gives marginally exact random draws with extremely low rejection rates \citep[see][]{PolsonScottWindle2013} for our dynamic logistic model.
Gradient‑based methods such as Hamiltonian Monte Carlo \citep{Neal2011} can also be used, but they require proposals on high-dimensional latent spaces, which is only practical in lower-dimensional setups.
Probit formulations exist \citep{AlbertChib1993, HeldHolmes2006}, but our model is logistic;
P\'{o}lya--Gamma augmentation is the natural, simple, and most practical choice.}

\section{Multivariate model}

We now extend the model to incorporate cross-series dependence.
We introduce a new index $k \in [K]$ that indexes the $k$-th time-series, where $K$ is fixed.
Let $y_{t,k}$ represent the $k$-th observable time-series at time $t$,
and $\bm{y}_t := (y_{t,1}, ..., y_{t,K})^\tr$,
and query the first $t'$ observations with $\bm{y}_{1:t'}$.
Similarly, write the $k$-th
latent trend,
measurement log-volatility,
and zero-probability
indexed at time $t$ as
$\theta_{t,k}$,
$h_{t,k}$,
and $p_{t,k} := \mathbb{P}(y_{t,k} = 0 \cond \pi_{t,k})$.
These are collected as $\bm{\theta}_t := (\theta_{t,1}, \ldots, \theta_{t,K})^\tr$, $\bm{h}_t := (h_{t,1}, \ldots, h_{t,K})^\tr$, and $\bm{\pi}_t := (\pi_{t,k}, \ldots, \pi_{t,K})^\tr$.

The multivariate zero-inflated UCSV (Z-MUCSV) model is specified as follows.
\begin{align*}
    y_{t,k} \cond p_{t,k}, y_{t,k}^*
        &\ind (1 - p_{t,k}) \, \delta_{y_{t,k}^*} + p_{t,k} \, \delta_0, \tag{$k \in [K]$} \\
    \bm{y}_t^* \cond \bm{\theta}_t, \bm{\Sigma}_t^{(y)}, \bm{h}_{t}
        &\ind \textsc{N}_K(\bm{\theta}_t, \bm{\Sigma}^{(y)}(\bm{h}_{t})), \\
    \bm{\theta}_t \cond \bm{\theta}_{t-1}, \sigma_{(\theta),1}^2, ..., \sigma_{(\theta),K}^2
        &\ind \textsc{N}_K(\bm{\theta}_{t-1}, \operatorname{diag}(\sigma_{(\theta),1}^2, ..., \sigma_{(\theta),K}^2)), \\
    \bm{h}_t \cond \bm{h}_{t-1}, \sigma_{(h),1}^2, ..., \sigma_{(h),K}^2
        &\ind \textsc{N}_K(\bm{h}_{t-1}, \operatorname{diag}(\sigma_{(h),1}^2, ..., \sigma_{(h),K}^2)), \\
    p_{t,k}
        &= \operatorname{logit}^{-1}(\pi_{t,k}), \tag{$k \in [K]$} \\
    \bm{\pi}_t \cond \bm{\pi}_{t-1}, \bm{\Sigma}^{(\pi)}
        &\ind \textsc{N}_K(\bm{\pi}_{t-1}, \bm{\Sigma}^{(\pi)}).
\end{align*}
We specify priors by
$\theta_{0,k} \ind \textsc{N}_1(m^{(\theta)}_k, v^{(\theta)}_k)$,
$h_{0,k} \ind \textsc{N}_1(m^{(h)}_k, v^{(h)}_k)$,
$\pi_{0,k} \ind \textsc{N}_1(m^{(\pi)}_k, v^{(\pi)}_k)$,
$\sigma_{(\theta),k}^2 \ind \textsc{IG}(\alpha_{k}^{(\theta)}, \beta_{k}^{(\theta)})$,
and $\sigma_{(h),k}^2 \ind \textsc{IG}(\alpha_{k}^{(h)}, \beta_{k}^{(h)})$
over $k\in[K]$.
Similarly to the univariate case, $\bm{y}_t^*$ is not necessarily assumed to be observed;
$y_{t,k} = y_{t,k}^*$ need not hold.
{To introduce dependencies, we first let $\bm{\Sigma}^{(\pi)} \sim \textsc{IW}(\nu^{(\pi)}, \bm{S}^{(\pi)})$,
where $\textsc{IW}(\nu, \bm{S})$ is an inverse Wishart distribution with degrees of freedom $\nu > K-1$ and $\bm{S}$ a $K \times K$ positive definite scale matrix.}

\subsection{Inter-component dependence with order-invariance}


A common approach to specifying inter-component dependence while preserving the time-varying volatility structure is through the Cholesky SV parameterization,
with
\begin{equation*}
    \bm{\Sigma}^{(y)}(\bm{h}_t) = \bm{C}^{-1}\operatorname{diag}(\exp{h_{t,1}}, ..., \exp{h_{t,K}})(\bm{C}^{-1})^\tr,
\end{equation*}
where $\bm{C}^{-1}$ is a strictly lower-triangular matrix of size $K \times K$ \citep{Primiceri2005, EoEzedaWong2023}.
This parameterization induces a normal regression-type representation under normal priors, which allows for inter‑component dependencies while conveniently facilitating tractable posterior sampling schemes.


However, the above parameterization depends on the ordering of variables.
This is critical in our setup, where the number of components is at least 49 (Japan);
assessing the sensitivity of posterior estimates to different orderings would be computationally intractable due to combinatorial explosion of possible permutations.
We therefore adopt an order-invariant parameterization of SV \citep{ChanKoopYu2024} by allowing the matrix $\bm{C}$ to remain non-singular but otherwise unrestricted.
Writing the $k$-th row of $\bm{C}$ as $\bm{C}_{k,:}$, let $\bm{C}$ be dense and not necessarily lower-triangular,
\begin{equation*}
    \bm{C}_{k,:} \ind \textsc{N}_K(\bm{m}^{(C)}_k, \bm{V}^{(C)}_k).
\end{equation*}
Alternative approaches to addressing the dependence on ordering include
\cite{LevyLopes2024} (and relatedly \citealp{Fruhwirth-SchnatterHosszejniLopes2024}) who also perform posterior inference on the series ordering.
Although applicable, the order‑invariant approach suffices for our purposes, as our primary goal is not to infer ordering but to identify co‑variation in an order‑invariant manner.

\subsection{Posterior inference}

The collection of unobserved parameters of interest in the multivariate case is
\begin{align*}
    \bm{\Theta}
    &:= \bracket{
        \theta_{0,1:K}, \;
        h_{0,1:K}, \;
        \pi_{0,1:K}, \;
        \bm{h}_{1:T}, \;
        \bm{C}, \;
        \sigma^2_{(\theta), 1:K}, \;
        \sigma^2_{(h), 1:K}, \;
        \bm{\Sigma}^{(\pi)}, \;
       \bm{\theta}_{1:T}, \;
        \bm{y}_{1:T}^*
    }.
\end{align*}
The joint distribution of the Z-MUCSV model is
\begin{align*}
    & 
    p(\bm{\Sigma}^{(\pi)})
    \bracket{
        \prod_{k=1}^K
            p(\bm{C}_{k,:}) \,
            p(\sigma_{(\theta),k}^2) \,
            p(\sigma_{(h),k}^2) \,
            p(\theta_{0,k}) \,
            p(h_{0,k}) \,
            p(\pi_{0,k}) \,
    } \\
    & \prod_{t=1}^T
        p(\bm{\theta}_t \cond \bm{\theta}_{t-1}, \sigma_{(\theta),1:K}^2) \,
        p(\bm{h}_t \cond \bm{h}_{t-1}, \sigma^2_{(h),1:K}) \,
        p(\bm{y}_t^* \cond \bm{\theta}_t, \bm{h}_t, \bm{C}) \,
        p(\bm{\pi}_t \cond \bm{\pi}_{t-1}, \bm{\Sigma}^{(\pi)}) \,
        \mathbb{P}(\bm{y}_t \cond \bm{y}_t^*, \bm{\pi}_t).
\end{align*}
Posterior sampling in the Z-MUCSV model is performed in blocks as in Algorithm \ref{alg:2}.
A detailed exposition is provided in the appendix.

\begin{center}
\spacingset{1}
\begin{minipage}{\linewidth} 
\begin{algorithm}[H]
    \medskip
    
\begin{enumerate}[(1)]
    \item Sample the initial states $(\bm{\theta}_0, \bm{h}_0, \bm{\pi}_0)$ given $(\bm{\theta}_1, \bm{h}_1, \bm{\pi}_1, \sigma^2_{(\theta),1:K}, \sigma^2_{(h),1:K}, \bm{\Sigma}^{(\pi)})$.
    
    \item Sample the static parameters $(\sigma^2_{(\theta),1:K}, \sigma^2_{(h),1:K}, \bm{\Sigma}^{(\pi)})$ given $(\bm{\theta}_{0:T}, \bm{h}_{0:T}, \bm{\pi}_{0:T})$.

    \item Sample the latent trend $\bm{\theta}_{1:T}$ given $(\bm{\theta}_0, \sigma^2_{(\theta),1:K}, \bm{C}, \bm{y}_{1:T}^*)$,
    
    \item Sample the stochastic volatility $\bm{h}_{1:T}$ given $(\bm{h}_0, \sigma^2_{(h),1:K}, \bm{C}, \bm{\theta}_{0:T}, \bm{y}_{1:T}^*)$.
    
    \item Sample the dynamic probabilities $\bm{p}_{1:T}$ given $(\bm{\pi}_0, \bm{\Sigma}^{(\pi)}, \bm{y}_{1:T})$.

    \item Data-augment non-zero $\bm{y}_{1:T}^*$ given $(\bm{\theta}_{1:T}, \bm{h}_{1:T}, \bm{C}, \bm{y}_{1:T})$.

    \item Sample $\bm{C}$ given $(\bm{y}_{1:T}^*, \bm{\theta}_{0:T}, \bm{h}_{1:T})$.
\end{enumerate}

    
    \caption{\texttt{Gibbs sampler: multivariate model}} 
    \label{alg:2}   
\end{algorithm}
\end{minipage}
\end{center}


\section{Results} \label{section:4}

\subsection{{Data and setup}} \label{section:4:data}

{Before presenting our results, we give a detailed description of the data beyond the overview in Section \ref{section:2}.
While an international standard for CPI measurement and calculation exists (e.g., by the International Labour Organization and promoted by the International Monetary Fund), important differences remain.
We describe our approach to best maintain consistency across countries in data selection and processing.}

\paragraph{Price index.}

{Table \ref{tab:data} is a summary of the data we use.
Our analysis utilizes quarterly disaggregated CPI data from four advanced economies: the US, UK, Germany, and Japan.
All data were obtained from the official statistical agencies of each country.
For the US, we use the CPI-U which covers all urban consumers (and is more comprehensive than the alternative CPI-W which is limited to urban wage earners and clerical workers).
For Germany, we use the COICOP-based CPI as this provided the most detailed disaggregated components.
Note that, although alternative indices exist (e.g., Personal Consumption Expenditures index in the US and the Retail Price Index in the UK), we use the CPI to best facilitate cross-country comparability.
}

\begin{table}[!h]
    \centering
    \spacingset{1.0}
    \begin{tabular}{p{1.9cm} p{1.7cm} p{1.5cm} p{2.9cm} p{3.9cm} p{.6cm}}
        \toprule
        \textbf{Source} & \textbf{Series} & \textbf{Transf.} & \textbf{Hierarchy} & \textbf{(e.g.)} & \textbf{$K$} \\
        \midrule
        US: BLS & Monthly CPI-U & QoQ & All items  
        
        $\supset$ Major group  
        
        $\supset$ \textbf{Class}
        
        $\supset$ Item stratum
        & \textit{All items}

        \textit{Food \& beverages}
        
        \textit{Fresh fruits}
        
        \textit{Apples}
        & 70
        \\ \midrule
        
        UK: ONS & Monthly CPI & QoQ & All items  
        
        $\supset$ Division

        $\supset$ Group
        
        $\supset$ \textbf{Class}
        
        $\supset$ Subclass
        & \textit{All items}
        
        \textit{Food \& non-alc.~bev.}
        
        \textit{Food}
        
        \textit{Fruit}
        
        \textit{Fresh or chilled fruit}
        & 76
        \\ \midrule
        
        Germany: Destatis & Monthly COICOP CPI & QoQ & All items
        
        $\supset$ Division
        
        $\supset$ Group
        
        $\supset$ \textbf{Class}
        
        $\supset$ Subclass
        & \textit{All items}
        
        \textit{Food \& non-alc.~bev.}
        
        \textit{Food}

        \textit{Fruit}
        
        \textit{Fresh or chilled fruit}
        & 85
        \\ \midrule
        
        Japan: SBJ & Monthly CPI & QoQ & All items
        
        $\supset$ Major group
        
        $\supset$ \textbf{Subgroup}
        
        $\supset$ Minor group
        & \textit{All items}
        
        \textit{Food}
        
        \textit{Fruit}

        \textit{Fresh fruit}
        & 49
        \\ \bottomrule
    \end{tabular}
    \spacingset{1.0}
    \caption{
        Summary of CPI data sources and disaggregation levels across countries.
        Hierarchy used in the analysis is \textbf{bolded}; lower levels exist and are acknowledged in Section \ref{section:4:data} with details on related considerations of data availability and structure.
        $K$ is the modeled number of disaggregated components.
        \textbf{Abbreviations}:
            BLS (Bureau of Labor Statistics);
            ONS (Office for National Statistics);
            Destatis; (Federal Statistical Office of Germany);
            SBJ (Statistics Bureau of Japan);
            CPI-U (CPI for all urban consumers);
            COICOP (Classification of Individual Consumption According to Purpose).
    }
    \label{tab:data}
\end{table}

{All data are at quarterly frequency, aligning with the convention in the majority of related literature on Bayesian trend-inflation estimation \citep[e.g.,][]{StockWatson2007, Chan2013, FaustWright2013, StockWatson2016, Chan2017, LiKoopman2021}.
Quarterly values are taken either directly from pre-computed three-month percentage changes or computed manually from monthly level data as Quarter-over-Quarter (QoQ) inflation rates.
For all countries, we utilize seasonally unadjusted data;
this is to best preserve the integrity of the \textit{raw} observations, as seasonal adjustment techniques typically involve predictive modeling, which may impose smoothing that can obscure the critical no-change signals we seek in disaggregated datasets.
Finally, we have uniformly aligned the data's time span to covering the period starting from 1988:Q2 to the latest available for all countries, as visible in Figure \ref{fig:intro:data2}.}

\paragraph{Disaggregation levels.}

{We employ the most granular disaggregation level for which expenditure weights are available, while ensuring that these levels are semantically consistent across countries.
The availability of weights is crucial for our subsequent aggregation exercise (Section \ref{sec:5:4}).
The specific levels are also summarized in Table \ref{tab:data}.
The data is not at the finest, rawest possible level of disaggregation since the weights were unavailable at such levels,
but it remains highly granular and captures the presence of zero-inflation, as visualized in Figure \ref{fig:intro:data2}.
We have intentionally excluded components which complicate the interpretation of correlations, or those unlikely to be contaminated with zero-inflation due to measurement error by construction, such as those labeled \dquote{not elsewhere classified} (nec) where their \textit{residual} definitions are inconsistent with other components.
This also improves computational stability and enhances the procedural robustness of our results.
}

\paragraph{Setup.}

In both the in- and out-of-sample settings, the prior hyperparameters are set as
$(m^{(\theta)}_{k}, v^{(\theta)}_{k}) = (m^{(h)}_k, v^{(h)}_k) = (0,10)$,
$(m^{(\pi)}_k, v^{(\pi)}_k) = (0, 1)$,
$(\alpha^{(\theta)}_{k}, \beta^{(\theta)}_{k}) = (\alpha^{(\pi)}_{k}, \beta^{(\pi)}_{k}) = (11, 1)$, and
$(\alpha^{(h)}_k, \beta^{(h)}_k) = (31, 1)$.
In place of $(\alpha^{(\pi)}_{k}, \beta^{(\pi)}_{k})$, Z-MUCSV is granted $(\nu^{(\pi)}, \bm{S}^{(\pi)}) = (2K, \bm{I}_K)$ and $(\bm{m}_k^{(C)}, \bm{V}_k^{(C)}) = (\bm{0}_K, \bm{I}_K)$.
We simulate $R=5000+1000$ MCMC samples and discard the initial $1000$ samples.
Data is scaled such that each component has equal standard deviations to facilitate stable posterior computation.

\subsection{Full in-sample estimation}

\subsubsection{Probability of zeros}

\begin{figure}[!h]
    \centering
    \subfloat[UCSV]{\vspace{4.0cm} \includegraphics[width=0.49\textwidth]{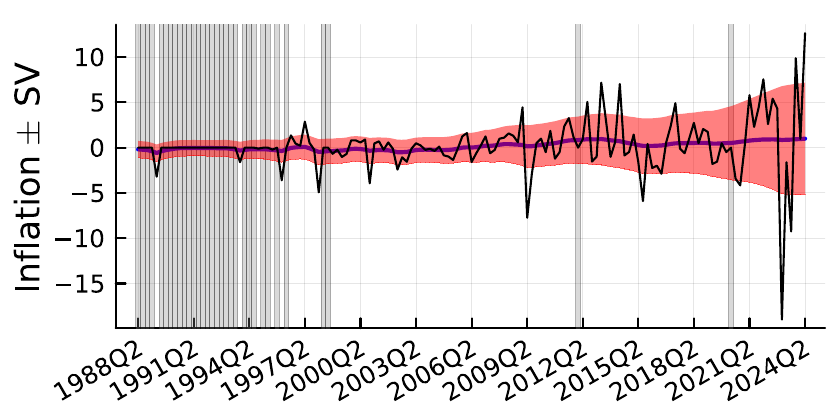}} ~
    \subfloat[Z-UCSV]{\includegraphics[width=0.49\textwidth]{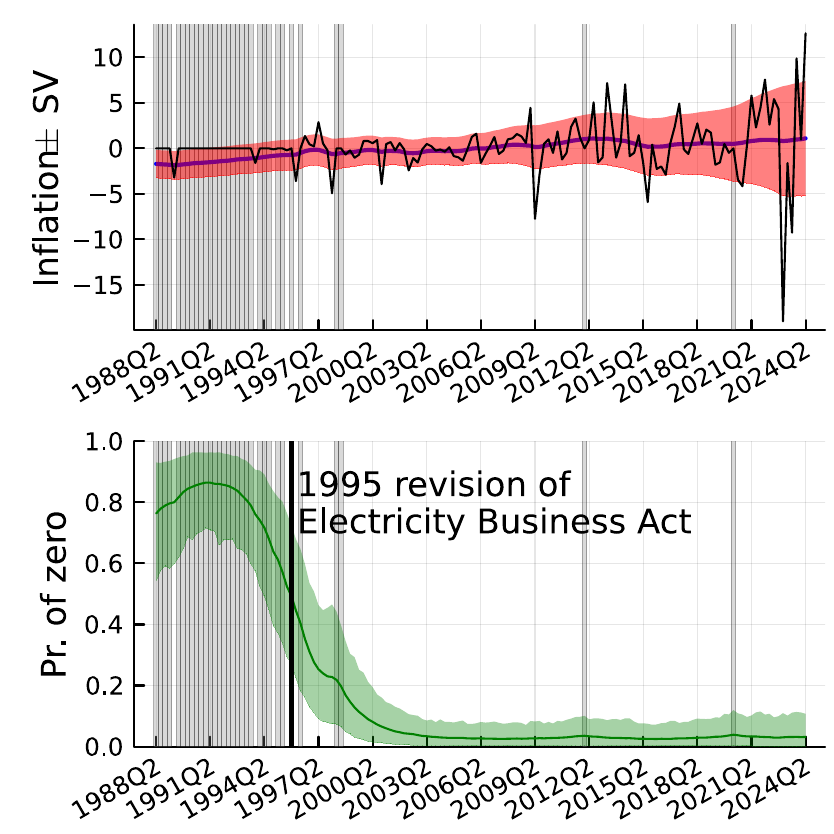}}
    \spacingset{1.1} \caption{
        \textbf{Top}: Time-series of \textit{electricity} inflation in Japan.
        \textcolor{blue}{\textbf{Solid blue line}}: posterior mean estimate of the trend.
        \textcolor{red}{\textbf{Red shaded region}}: posterior mean estimate of the time-varying standard deviation around the trend, $\pm\exp(h_t/2)$; stochastic volatility estimate.
        \textbf{Bottom}, \textcolor{Green}{\textbf{solid green line}}: posterior mean estimate of time-varying probability of zero.
        \textcolor{Green}{\textbf{Green shaded region}}: 90\% credible interval.
    }
    \label{fig:results:prob}
\end{figure}

We begin by presenting posterior estimates.
Figure \ref{fig:results:prob} first shows the major distinction between (M)UCSV and Z-(M)UCSV in that the latter assigns non-zero time- and component-specific probabilities of zeros.
Z-UCSV quantifies the significant heterogeneity in price staleness during the periods before and after December 1995 (1995:Q4), showing a rapidly declining time-specific probability of zeros.
This period corresponds to the 1995 revision of the Electricity Business Act, which was a pivotal policy initiative for energy liberalization (i.e., price deregulation) in Japan.

\subsubsection{Trend-inflation}

\paragraph{Interpretation.}

{
It is important to note that the interpretation of the latent trend component differs structurally between zero-inflated and non-zero-inflated models.
In the zero-inflated model, the trend corresponds to the expected value {\em conditional} on non-zero observations, whereas the non-zero-inflated model estimates are {\em unconditional}.
The implication is that one should conceptually distinguish {\em expected} inflation (in the observable-data space) from {\em trend} inflation (which is latent), as the two coincide approximately in the standard non-zero‑inflated model, but they need not do so in the zero‑inflated setting.
When zeros are negligible, both models yield similar posterior distributions and thus trend inflation estimates.
When non-negligible, the model, by design, decomposes the observed inflation rate into a sparse realization and a latent trend component;
the latter enables additional inference on the latent and conditional trend.
}

\paragraph{Comparison.}

\begin{figure}[!h]
    \spacingset{1.1} \caption{
        Inflation time-series of \textit{games of chance} in Germany,
        and posterior estimates of the
        trend (\textcolor{blue}{\textbf{solid blue line}}),
        predictive mean (\textbf{dotted black line}),
        marginal time-varying volatility in standard deviation (\textcolor{red}{\textbf{solid red line}}), and
        probability of zero (\textcolor{Green}{\textbf{solid green line}}).
        \textbf{Shaded regions}: 90\% credible interval.
    }
    \label{fig:results:trend:1}
    \centering
    \subfloat[UCSV]{\includegraphics[width=0.49\textwidth]{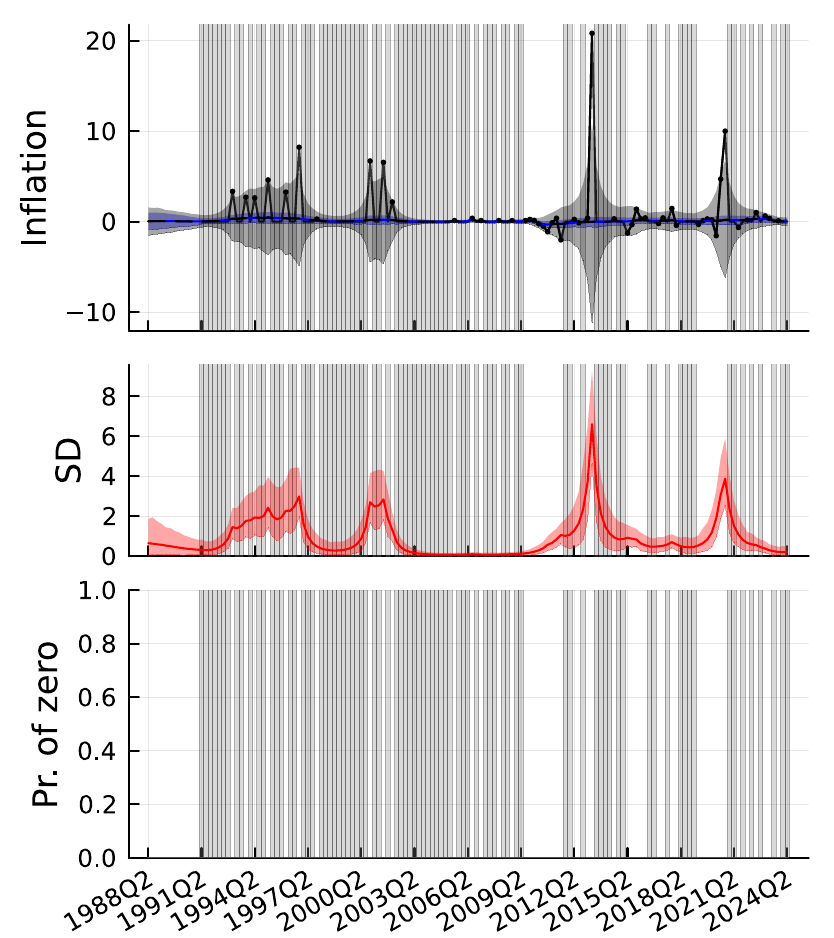} \label{fig:results:trend:1:a}} ~
    \subfloat[Z-UCSV]{\includegraphics[width=0.49\textwidth]{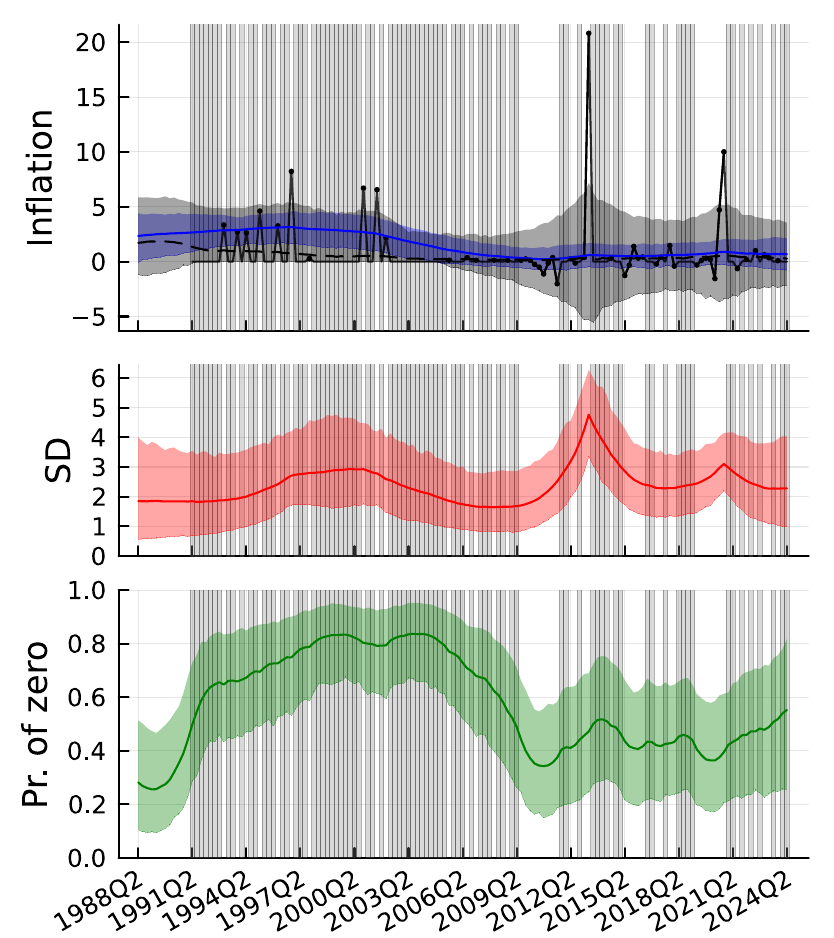} \label{fig:results:trend:1:b}} \\ ~ \\ ~ \\
    \subfloat[MUCSV]{\includegraphics[width=0.49\textwidth]{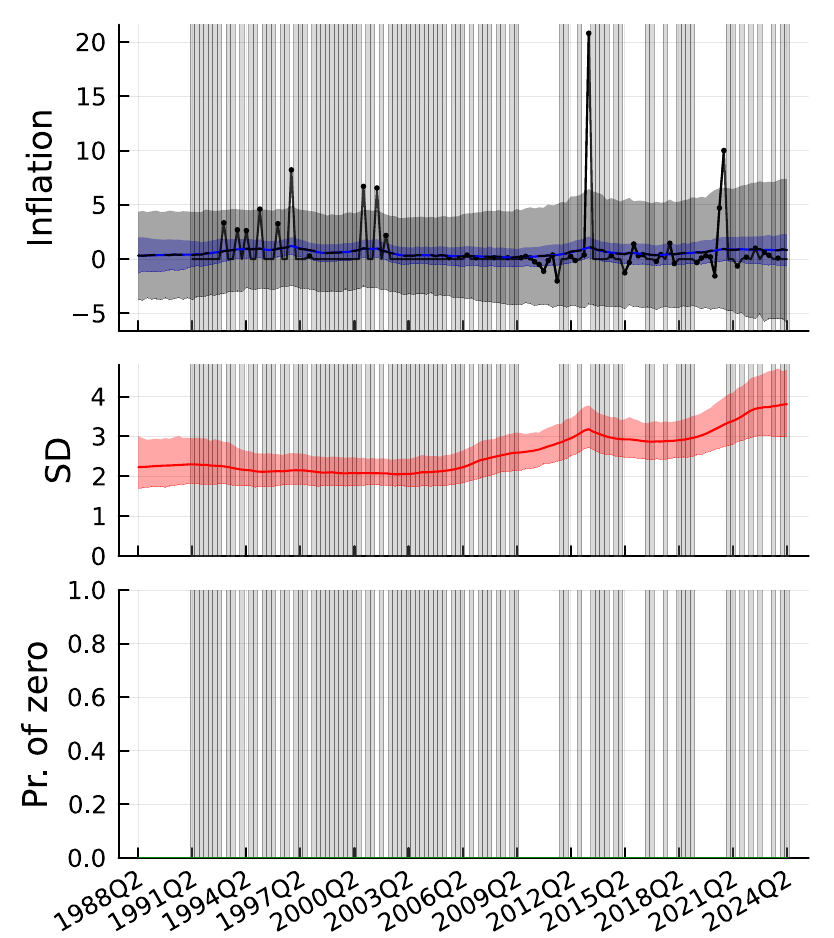} \label{fig:results:trend:1:c}} ~
    \subfloat[Z-MUCSV]{\includegraphics[width=0.49\textwidth]{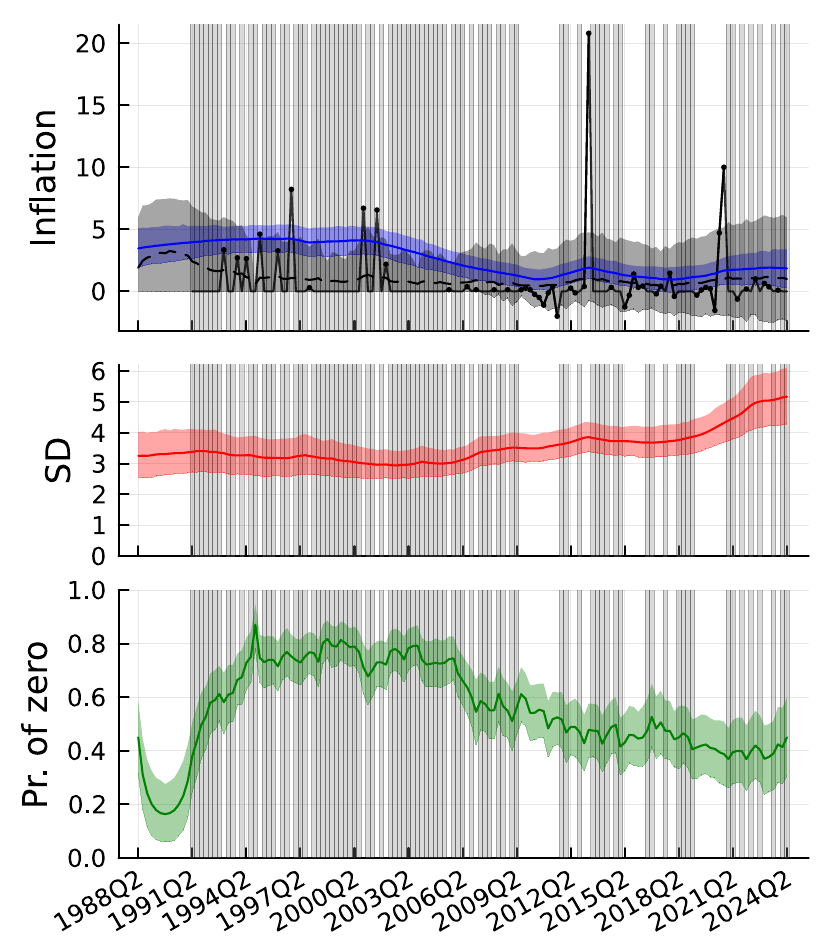} \label{fig:results:trend:1:d}}
\end{figure}

Augmenting the UCSV model with a time-varying probability of zeros enables recovery of more informative trend fluctuations.
For example, in Figure \ref{fig:results:trend:1}, zeros are extremely frequent and the observables are nearly exclusively non-negative, which suggests an underlying process that cannot be adequately or sensibly described by smoothly evolving trends with standard symmetric noise.
The non-zero-inflated UCSV model gives a mere dynamic summary of the central tendency but is inevitably drawn toward zero in the presence of repeated zeros, resulting in latent trend and volatility estimates that are confidently conservative and uninformative;
both the expected and trend inflation from the non-zero‑inflated model remain close to 0--1\% (i.e., Figures \ref{fig:results:trend:1:a} \& \ref{fig:results:trend:1:c}).
In contrast, the trend inflation from the zero‑inflated model rises upto 5\%, above the 2\% threshold, for this component in interpolating the conditional signals and better align with the available non‑zero observations.
This is arguably relevant for such components with infrequent adjustments, where prices are often governed by administrative rules or long-term contracts.
Rather than treating extended periods of zero-inflation as definitive evidence of a near-zero underlying trend, which conveys limited information, the model recovers latent inflationary pressure that may persist despite the absence of observed price changes.
Further plots are given in the appendix.

\begin{figure}[h]
    \centering
    \includegraphics[width=\textwidth]{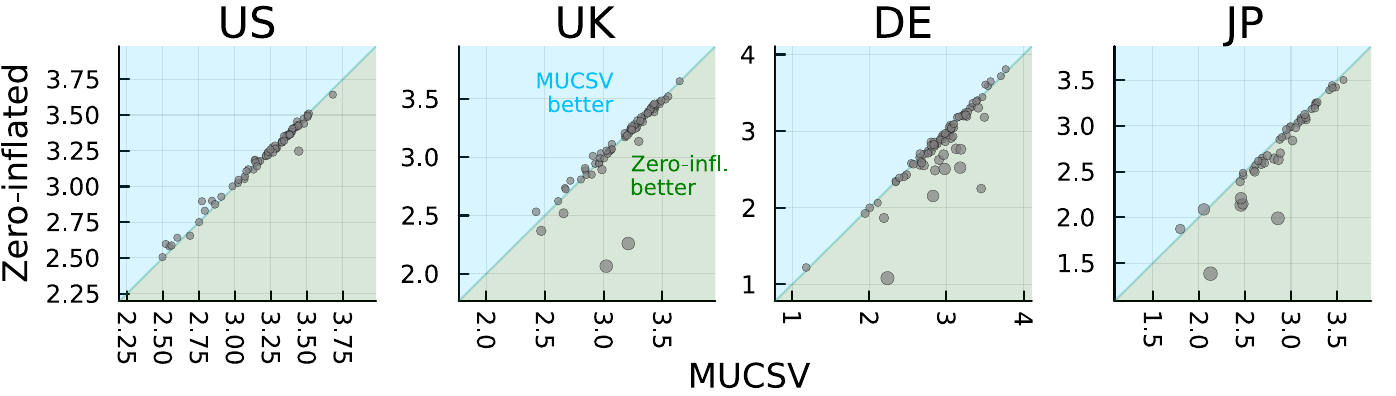}
    \spacingset{1.1} \caption{
        Comparison of in‑sample fit between the \textcolor{blue}{\textbf{standard multivariate model}} and the \textcolor{Green}{\textbf{zero‑inflated model}}.
        Each point represents the mean absolute error (MAE) of the posterior predictive median for a single component.
        The transparent red region indicates where the zero‑inflated model yields in-sample improvements.
        Point size is proportional to the proportion of observed zeros.
    }
    \label{fig:results:in-sample}
\end{figure}

The improved in‑sample fit across components is also evident in Figure \ref{fig:results:in-sample}.
Components with minimal or negligible zero-inflation cluster near the 45-degree line, which indicates comparable performance between the two models.
Then, in contrast, components with higher levels of zero‑inflation deviate from this line, and consistently favors the zero‑inflated specification.
The pattern supports that the zero‑inflation \textit{add-on} provides meaningful benefits precisely in settings where zero‑inflation is substantial.

\subsubsection{Volatility estimation} \label{sec:results:vol}

Posterior estimates of time-varying log-volatilities are essential for assessing whether observed variations are permanent or transitory.
With persistent zeros, volatilities may not necessarily be accurately reflected in the data, and the model should account for the limited information on volatility (see Section \ref{section:2}).
Figure \ref{fig:results:trend:1} illustrates this;
compared with UCSVs, Z‑UCSVs are far less susceptible to irregular spikes caused by intermittent exact zeros in the data, and instead estimate higher and smoother regularized volatility by interpolating these periods in which only zeros are observed.
UCSVs produce confidently lower volatility estimates and are highly sensitive to zero/non‑zero fluctuations, because zeros are incorporated directly into the likelihood as if they were generated from a Gaussian measurement.

\begin{figure}[!h]
    \centering
    \includegraphics[width=\textwidth]{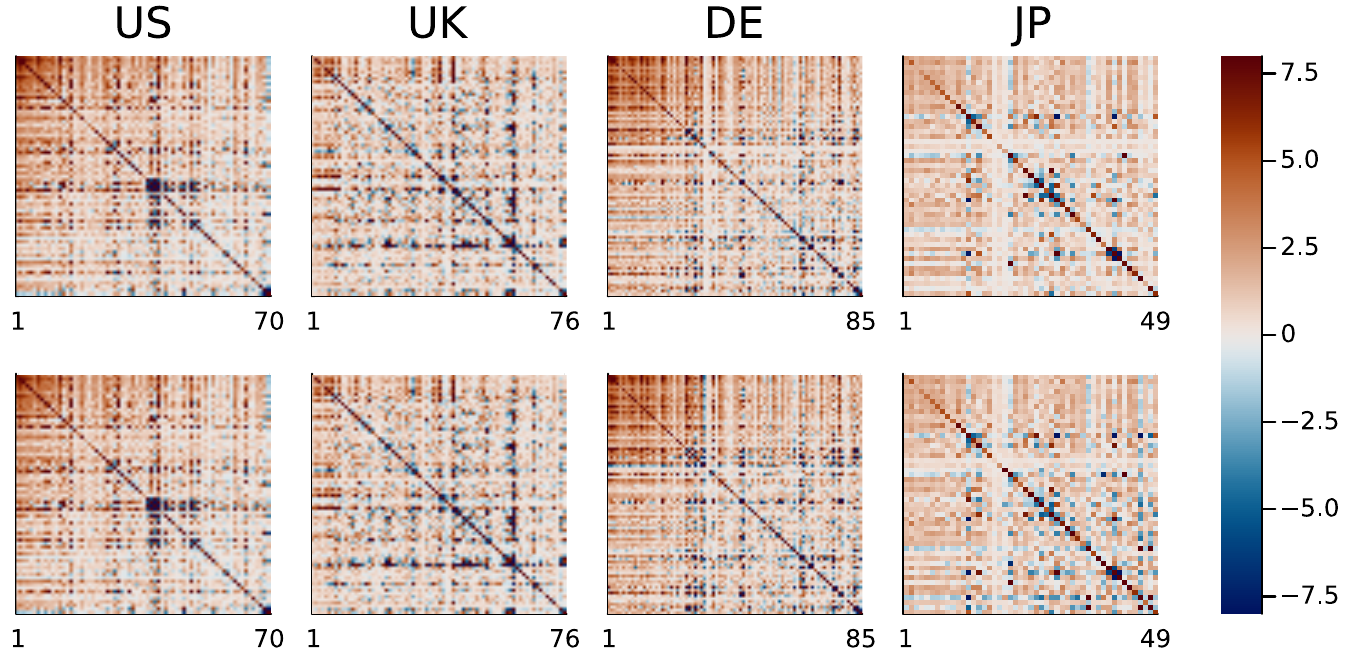}
    \spacingset{1.1}
    \caption{
        Heatmap of the posterior mean of the covariance $\bm{\Sigma}^{(y)}(\bm{h}_t)$ (fixed at the latest available time point $t=T$ for visibility) illustrating the contemporaneous covariation among nonzero components in each model.
        \textbf{Top}: standard MUCSV.
        \textbf{Bottom}: zero-inflated MUCSV.
    }
    \label{fig:results:cov}
\end{figure}

Figure \ref{fig:results:cov} illustrates the marginal posterior mean estimates for the covariance.
Albeit subtle, for Germany and Japan where zeros are more frequent, the non‑zero‑inflated model attenuates the negative covariances that the zero‑inflated model detects.

\subsection{Out-of-sample forecasting exercise} \label{section:5}

\subsubsection{Setup}

We present results from a recursive out-of-sample forecasting exercise conducted as follows.
We begin by designating a data window of the first 45 available quarters of post-processed data from 1988:Q2 to 1999:Q2 as the \textit{training} set, and the subsequent $H=8$ quarters starting from 1999:Q3 as the holdout \textit{validation} set.
Using the training set and its associated index set $[T]$, we approximate the $h$-quarter-ahead posterior predictive distribution via forwardly simulating the thinned MCMC samples of vector forecasts, $\bm{y}_{T+h | T}^{(r)}$ for $h \in [H]$ (with $r$ indexing the $r$-th thinned MCMC sample).

Both point and interval forecasts are considered.
As for point forecasts, we compute the posterior median of MCMC samples $(\bm{y}_{T+h | T}^{(r)})_{r=1}^R$, element-wise;
the median gives an appropriate and tractable summary of the posterior predictive distribution which assigns positive probability to exact zeros.
Interval forecast is approximated as a $(1-\alpha)$\% credible interval (CI), approximated also from MCMC draws.

Then, the next window of training samples with expanding window size (i.e., varying $T$) is used to compute the point forecast over the next holdout set.
We repeat this until $T$ is such that its $(H=8)$-quarters-ahead forecast corresponds to the final available data.
In the final window,
the training set ranges from 1988:Q2 to 2022:Q2, and
the holdout set ranges from 2022:Q3 to 2024:Q2.
Out-of-sample forecast performance is evaluated using
(a) the empirical mean absolute error (MAE) for point forecasts, relative to non-missing observations, and
(b) calibration accuracy of interval forecasts at different levels of $\alpha$.
{Note that log predictive density scores (LPDS), while standard for evaluating interval/density forecasts, are inapplicable for comparing UCSV versus Z-UCSV, as the latter's predictive distribution is partially a point-mass.}

\subsubsection{Point forecast}

\begin{figure}[h]
    \centering
    \includegraphics[width=\textwidth]{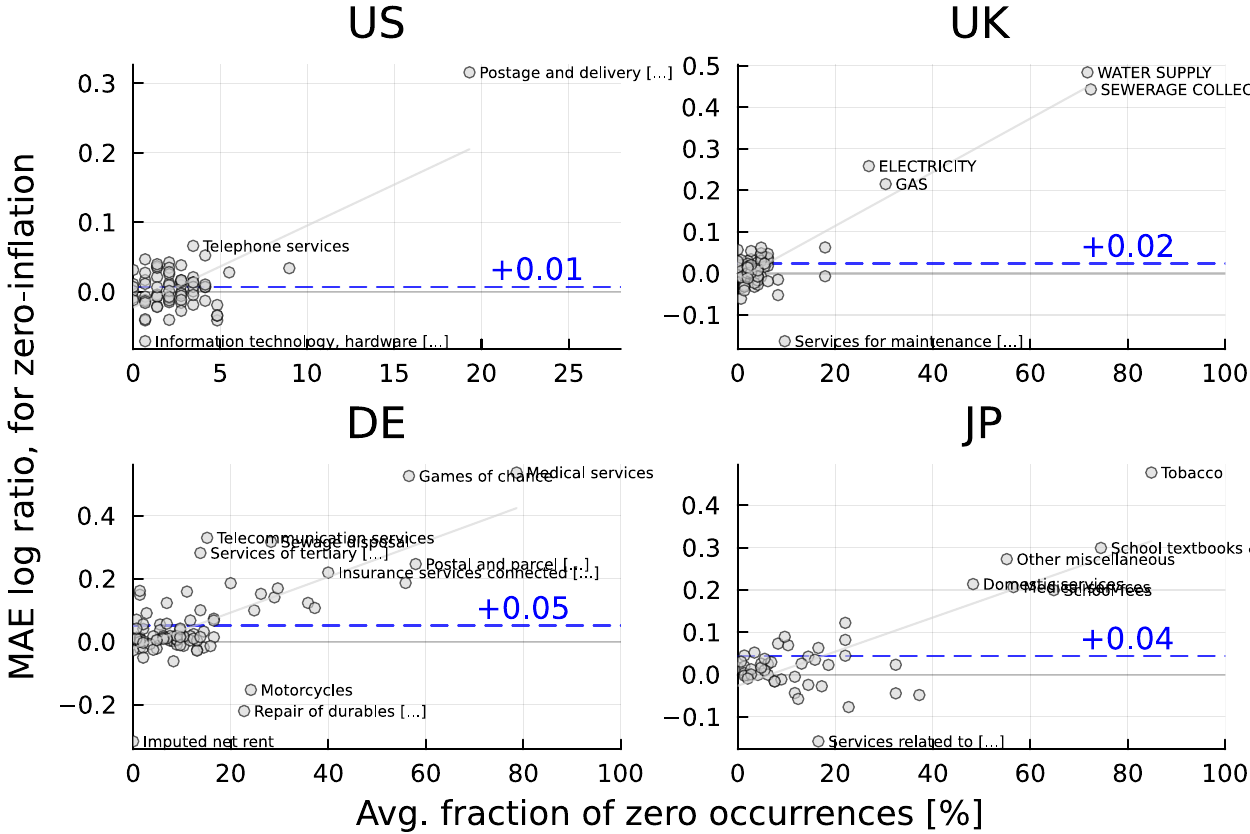}
    \spacingset{1.1} \caption{
        Component specific $(h=1)$-quarter-ahead point-forecast performance comparisons; higher values indicate that the point forecast was better for the zero-inflated model.
        \textcolor{blue}{\textbf{Blue dashed line}}: average over the components.
    }
    \label{fig:results:forecast:1}
\end{figure}

Figure \ref{fig:results:forecast:1} compares the one-quarter-ahead forecast MAEs for multivariate models.
Many components show comparable performance across models.
Important exceptions, however, are evident for some components where zero-inflation is relevant: most notably in the UK, Germany, and Japan.
The zero-inflated model
(a) provides improved point forecasts on average compared to the non-zero-inflated model, and
(b) the improvements appear to be associated with the extent to which exact zeros are prevalent.
These illustrate how the zero‑inflated specification provides value;
it delivers superior forecast performance when zero‑inflation is prevalent,
and effectively reverts to the non‑zero‑inflated specification (and performing comparably to the non‑zero‑inflated model) when zero‑inflation is limited.
{Further results (with similar patterns) for
longer horizons at eight quarters ahead,
univariate models without cross-series dependence,
and a different metric (mean squared error as loss) are relegated to the appendix to preserve space.}

\subsubsection{Interval forecast}

\begin{figure}
    \centering
    \includegraphics[width=\linewidth]{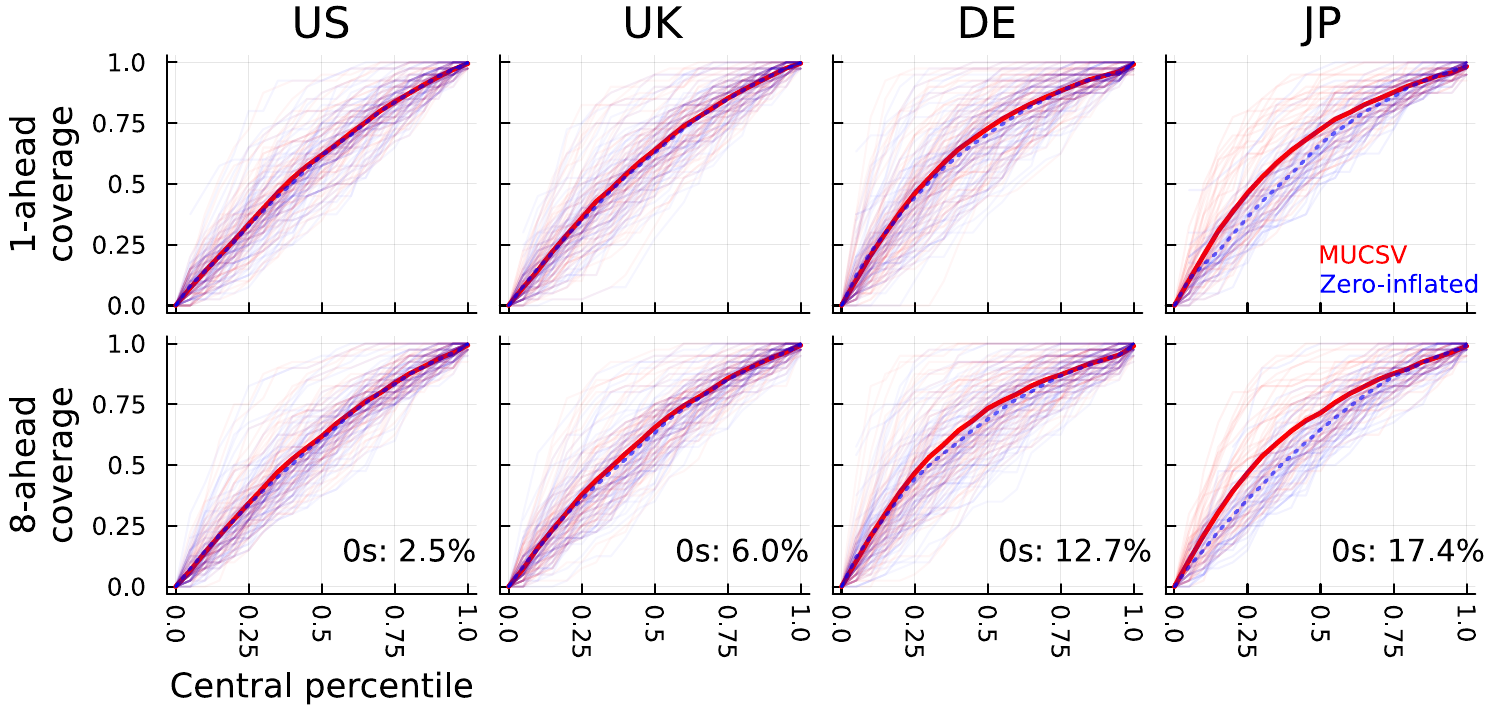}
    \spacingset{1.1} \caption{
        Empirical coverage of 1- and 8-quarter-ahead $100\alpha$\% interval forecasts at different levels of $\alpha \in (0,1)$ for the multivariate model; alignment to the 45-degree line (not shown) indicates perfect calibration.
        Average calibration performance is summarized by a \textcolor{red}{\textbf{red solid line (MUCSV: non-zero-inflated)}} and a \textcolor{blue}{\textbf{blue dotted line (ZMUCSV: zero-inflated)}}.
    }
    \label{fig:results:interval:all}
\end{figure}

Figure \ref{fig:results:interval:all} shows the empirical coverage rates of the interval forecasts for the multivariate models at various percentile levels.
Alignment with the 45-degree line (not shown) would indicate the well-calibrated ideal where the $(1-\alpha)$\% CI covers $\alpha$\% of the out-of-sample data.
Given the high number of components, we summarize by averaging the coverage rates at each level.
(We note that \textit{left-tails} for some calibration lines for the zero-inflated model is missing, as we have omitted regions where the prediction intervals have totally degenerated to $\set{0}$ and such regions only yield summaries on the data about the proportion of zeros in the hold-out observations which is potentially misleading.)
The zero-inflated model outperforms the non-zero-inflated model being on average closer to the 45-degree line: noticeably pronounced with Japan with the highest prevalence of zeros.

\begin{figure}[!h]
    \centering
    \subfloat[Transportation-related components]{\includegraphics[width=\textwidth]{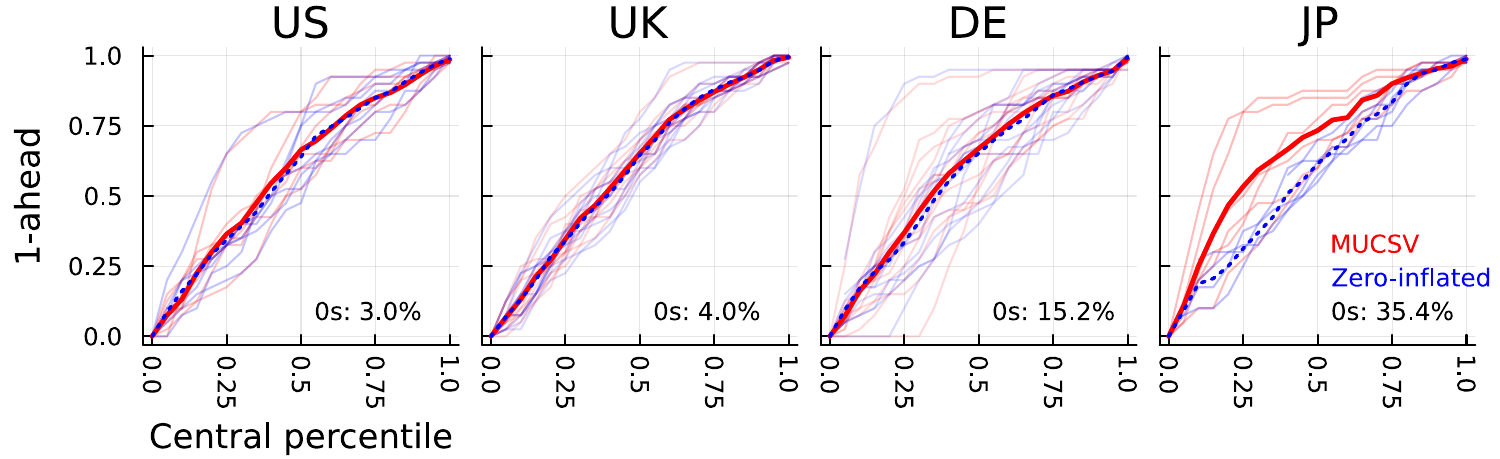}} \\
    \subfloat[Health-related components]{\includegraphics[width=\textwidth]{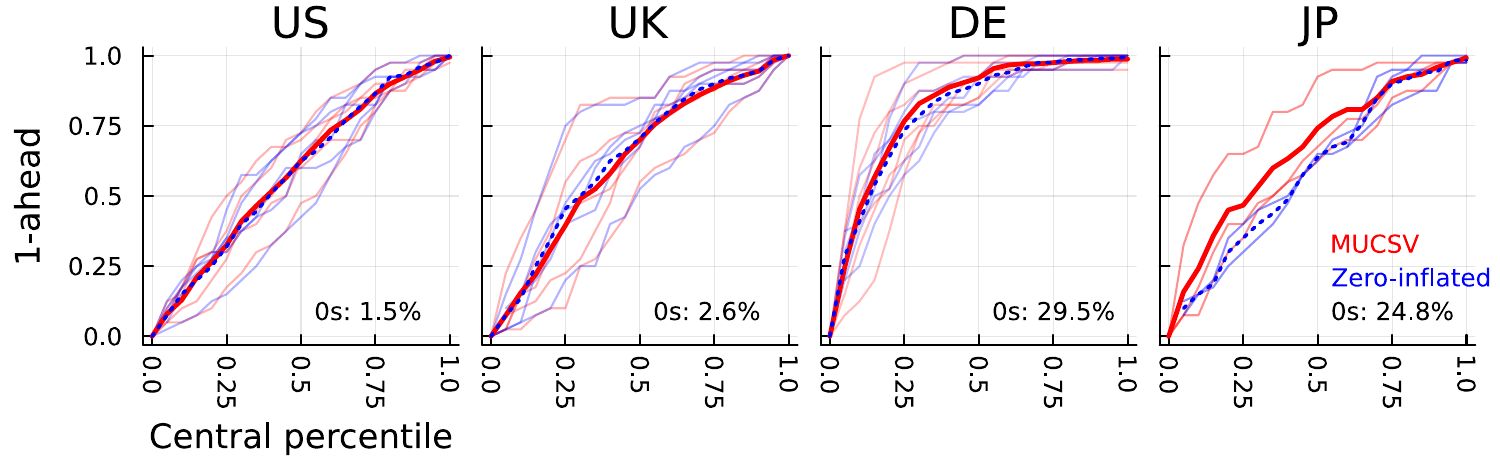}}
    \spacingset{1.1} \caption{
        Empirical coverage of 1-quarter-ahead $100\alpha$\% interval forecasts at different levels of $\alpha \in (0,1)$ for selected components; alignment to the 45-degree line (not shown) indicates perfect calibration.
        Average calibration performance is summarized by a \textcolor{red}{\textbf{red solid line (MUCSV: non-zero-inflated)}} and a \textcolor{blue}{\textbf{blue dotted line (ZMUCSV: zero-inflated)}}.
    }
    \label{fig:results:interval:sub:h=1}
\end{figure}

We further decompose the interval forecast results by \textit{major} divisions (i.e., COICOP level 2).
Figure \ref{fig:results:interval:sub:h=1} shows two categories that exhibit substantial observable differences in the averaged calibration lines:
transport- and health-related components.
These items tend to undergo price revisions only after administrative processes, resulting in intermittent observable inflation and thus persistent zeros.
These further support the use of the zero-inflated model, as it delivers better-calibrated interval forecasts on average where zeros are prevalent, while maintaining comparable performance to the non-zero-inflated model when zeros are infrequent.
Figures for other cases in the appendix also exhibit similar patterns favoring the zero-inflated model.

\newpage

\subsection{Aggregating the disaggregates} \label{sec:5:4}

{
A major appeal of the UCSV model is its ability to address the fundamental challenge in inflation measurement, of extracting a smoothed estimate of aggregate trend-inflation from noisy observations.
This gives a practical gauge of how closely inflation is anchored to the aggregate target.
It would therefore be informative to aggregate the individual component-wise estimates from the Z-UCSV model and validate relative to the overall measure.
This section presents both in-sample and out-of-sample results.
}

{Aggregation is feasible due to our design of the data collection outlined in Section \ref{section:4:data} where we ensured the availability of weights.
The process begins with estimating component-wise trends, which implicitly defines a posterior for the level of the \textit{denoised} price index.
These can then be hierarchically aggregated, then converted into trend-inflation at varying levels of disaggregation.
We compare this approach (\textit{component-wise Z-UCSV}) with the trend-inflation estimated from the standard UCSV model applied to aggregated CPI (\textit{direct UCSV}).
We note that the aggregated data used in the direct UCSV approach has been manually computed based on the latest available set of fixed items and weights at the time of analysis;
this is due to complications such as incomplete historical information on evolving weighting schemes, item introductions and/or discontinuations, and country-specific variations.}

\begin{figure}[!h]
    \centering
    \subfloat[]{\includegraphics[width=0.49\textwidth]{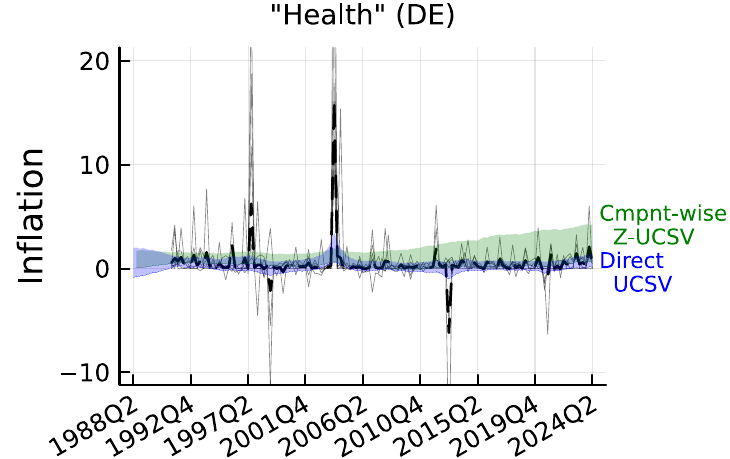} \label{fig:results:agg:in:3}} ~
    \subfloat[]{\includegraphics[width=0.49\textwidth]{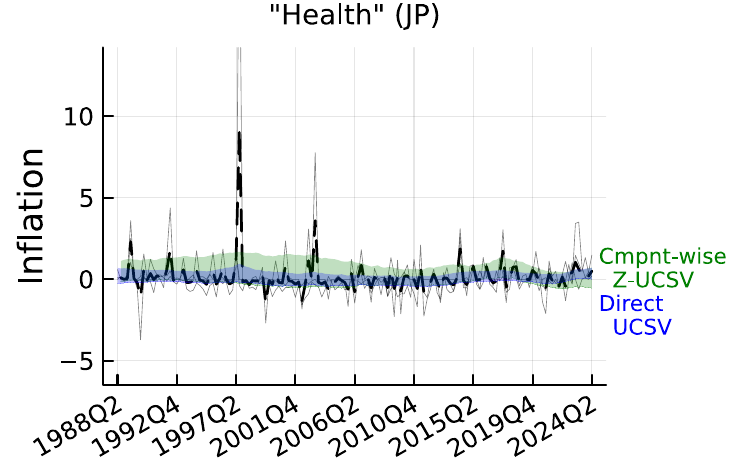} \label{fig:results:agg:in:4}} \\
    \subfloat[]{\includegraphics[width=0.49\textwidth]{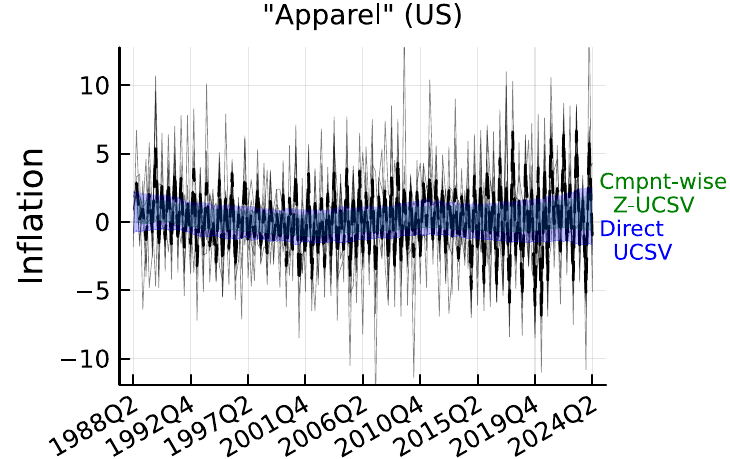} \label{fig:results:agg:in:1}} ~
    \subfloat[]{\includegraphics[width=0.49\textwidth]{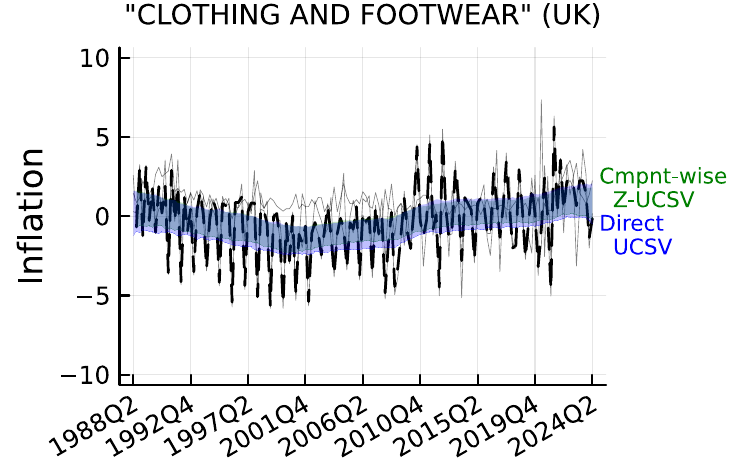} \label{fig:results:agg:in:2}} \\
    \spacingset{1.1} \caption{
        Selected visual comparison of implied trend‑inflation estimates from \textcolor{Green}{\textbf{component‑wise Z‑UCSV}} (disaggregate trends aggregated) and \textcolor{blue}{\textbf{direct UCSV}} (model fitted directly to the aggregate), shown separately for a major category.
        \textbf{Shaded regions} indicate the 75\% posterior credible interval.
        \textbf{Dashed black line}: aggregated index.
        \textbf{Solid thinner gray lines}: constituent indices.
    }
    \label{fig:results:agg:in}
\end{figure}

{Figure \ref{fig:results:agg:in} illustrates differences in inflation estimates that are non-negligible (e.g., Figures \ref{fig:results:agg:in:3} \& \ref{fig:results:agg:in:4}) and
similar (e.g., Figures \ref{fig:results:agg:in:1} \&; \ref{fig:results:agg:in:2}).
Similar qualitative pattern is consistent across categories.
The \textit{direct UCSV} approach is rather sensitive to short‑lived spikes in the aggregate index (e.g., Figure \ref{fig:results:agg:in:3}) and produces trend estimates that are highly conservative and often shrunken near zero.
The \textit{component‑wise Z‑UCSV} approach produces smoother and less overly conservative estimates that better reflect episodes of zero-inflation.
For categories with frequent price adjustments, both yield nearly identical trend estimates.
The disaggregate structure hence matters most in low‑adjustment environments where \textit{direct UCSV} produces uninformative fluctuations and \textit{component‑wise Z‑UCSV} automatically responds appropriately to for zero‑inflation episodes.}

\begin{figure}[!h]
    \centering
    \includegraphics[width=\textwidth]{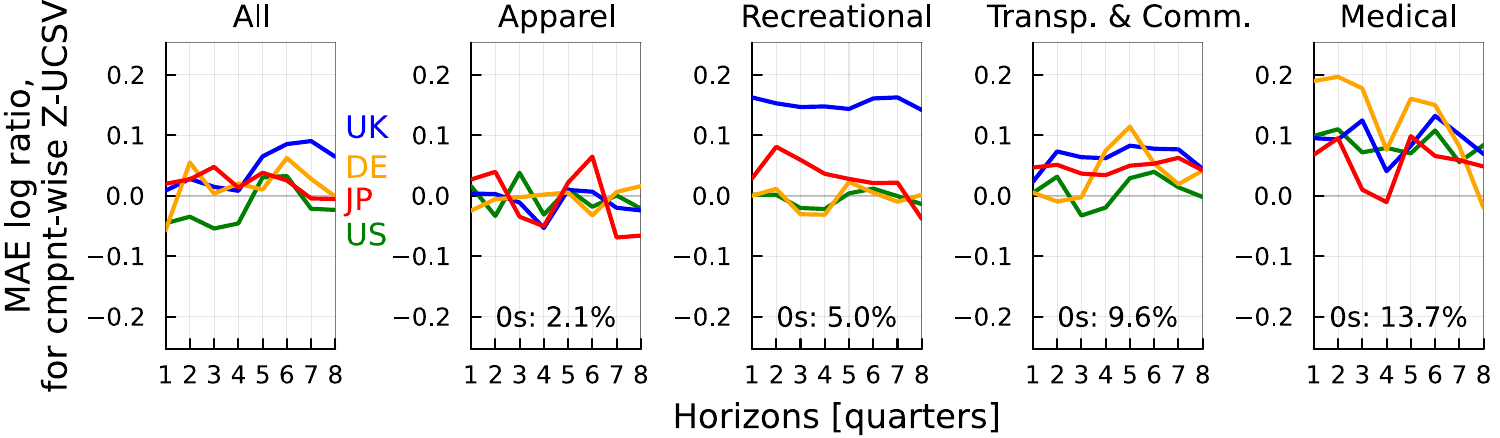}
    \spacingset{1.1} \caption{
        Out-of-sample forecast comparison between
        (a) \textit{component-wise Z-UCSV} (disaggregate trends aggregated) and (b) \textit{direct UCSV} (directly fitting the aggregated disaggregates), measured via MAE.
        Results are shown by overall index and select major groups.
        \textbf{Note}: Higher values favor \textit{component-wise Z-UCSV}.
    }
    \label{fig:results:agg:out}
\end{figure}

{
Figure \ref{fig:results:agg:out} summarizes the implications in an out‑of‑sample setting.
When forecasts are fully aggregated (leftmost panel), the zero‑inflated model performs similarly to the direct UCSV approach.
Major‑group level inspection then reveals a more heterogeneous pattern.
For groups where zero observations are relatively unimportant, such as apparel (as in Figures \ref{fig:results:agg:in:3} \& \ref{fig:results:agg:in:4}), the posteriors and thus the out‑of‑sample predictive performance remain similar.
For non‑negligible groups, such as health‑related major categories (as in Figures \ref{fig:results:agg:in:1} \& \ref{fig:results:agg:in:2}), the zero‑inflated specification gives clearer out‑of‑sample gains.
}

\subsection{Implications}

We end by summarizing the implications of (not) choosing a zero-inflated model in two substantive respects.
First, aggregated models that smooth away measurement errors in disaggregated data tend to understate underlying trend inflation when zeros are non‑negligible, whereas a zero‑inflated model gives informative and uncertainty-aware latent estimates that distinguish expected from trend inflation.
Such better reflects the risk‑averse stance in inflation measurement and decision‑making which involves asymmetric risks \citep{DoladoPedreroRuge-Murcia2004, Surico2007, KilianManganelli2008}, in that entities favor more sensitive detection over delayed recognition given that the cost of overlooking an inflationary episode outweighs the cost of issuing a false alarm.
For instance, in real environments involving inflation target bands \citep[e.g.,][$2 \pm 1\%$ band]{BoE2023}, relying on estimates that do not account for zero-inflation may understate infrequent, yet substantial, price adjustments such as within or via sectors where rising input costs are not immediately reflected in observable price changes (e.g., \ref{fig:results:agg:in:1} \& \ref{fig:results:agg:in:2}).
The same applies to volatility, where prolonged zeros lead non-zero-inflated models to produce confidently subdued estimates (see Section \ref{sec:results:vol}).

Second, the model-based approach does not rely on manual flagging.
Certain instances of price stickiness may be known in advance, but when tracking numerous component-disaggregates --- as in our application --- not all irregular price adjustments/staleness would be predictable \textit{ex ante}, and the zero-inflated model can be useful for detecting latent inflationary signals that may otherwise be overlooked.
For instance, transportation fees can be adjusted infrequently in response to infrastructure developments by semi-municipal/regional authorities (e.g., transit fares for buses, subways, and commuter rail).
Healthcare fees can be shaped by insurance schemes and subsidies in response to external variabilities such as budget cycles or subsidies (e.g., copayments and service fees).
The model accommodates for above such emphases.


\section{Concluding remarks and future research} \label{section:6}

We introduced a novel zero-inflated SV model that estimates the time-varying probability of exact zeros, upon the flexible framework of DGLMs.
A practical posterior Gibbs sampler utilizing P\'{o}lya--Gamma augmentation was developed to sample the time-varying probabilities efficiently.
Using a comprehensive dataset of disaggregated CPI inflation for advanced economies with varying degrees of zero‑inflation, we showed that the zero‑inflated specification captures more informative fluctuations in time‑varying trends and volatility than standard SV models which tended to produce overly conservative or even erroneous estimates due to their assumption that exact zeros are generated via Gaussian likelihoods, and the consequent sensitivity to inherent spikes in the data.
In forecasting, introducing zero-inflation improved point and interval forecast performance over the non-zero-inflated counterpart, particularly when zeros were prevalent.

We conclude by mentioning avenues for further research.
An unanswered substantive question is: to what extent price staleness contributes to aggregate inflation persistence \citep{CogleyPrimiceriSargent2010, ChanKoopPotter2016, HwuKim2019} and how the Z-(M)UCSV model might quantify this relationship.
An important methodological direction would be incorporating existing extensions of univariate (UC)SV models.
Deterministic volatility feedback mechanisms have been explored in univariate (UC)SV models, such as those in \cite{Chan2017} and \cite{HuberPfarrhofer2021}, through SV-in-mean models with time-varying parameters.
It may be worthwhile to incorporate such alongside our proposed zero-inflated specification for improved fit.

Another natural question is how the components co-move over time.
One modeling avenue for addressing such is dynamic latent factor models \citep{AguilarWest2000, LopesCarvalho2007, DelNegroOtrok2008, GruberWest2016BA, LavineCronWest2020factorDGLMs}, which are widely used in macroeconomic and financial time-series to capture interpretable co-movements in a parsimonious manner.
Incorporating such structures was beyond the scope of the present study, which focused specifically on why one should and how to model/infer zero-inflation by extending the standard UCSV model.
Factor structures can, in principle, be applied to any subset of the zero-inflated model's dynamic latent states (i.e., trend, volatility, or zero-inflation probability).
Doing so however would require substantial model‑selection or -mixing work, and a clear methodologically tractable procedure for guiding such extensions in the present context is not obvious.
The conceptual consideration and extension would merit an extensive study of its own, to expand our understanding of price inflation dynamics.

\if0\blind
{

\section*{Acknowledgments}
We thank Mototsugu Shintani, Jouchi Nakajima, and Yuko Onishi for their helpful comments on the early version of this paper. 

\section*{Funding}
IK's research was partly supported by JSPS KAKENHI Grant Number 22K13374 from Japan Society for the Promotion of Science.

\section*{Disclosure Statement}
The authors report that there are no competing interests to declare.

} \fi

\if1\blind
{
} \fi

\bibliographystyle{apalike}
\bibliography{ref.bib}

\end{document}